\documentclass[pdflatex,sn-mathphys-num]{sn-jnl}% Math and Physical Sciences Numbered Reference Style
\usepackage{graphics}

\usepackage{graphicx}%
\usepackage{multirow}%
\usepackage{amsmath,amssymb,amsfonts}%
\usepackage{amsthm}%
\usepackage{mathrsfs}%
\usepackage[title]{appendix}%
\usepackage{xcolor}%
\usepackage{textcomp}%
\usepackage{manyfoot}%
\usepackage{booktabs}%
\usepackage{algorithm}%
\usepackage{adjustbox}
\usepackage{pdflscape}
\usepackage{algorithmicx}%
\usepackage{algpseudocode}%
\usepackage{listings}%
\usepackage{bold-extra}
\usepackage[normalem]{ulem}

\newcommand{\cii}{[C\,{\scriptsize II}]\,}
\newcommand{\nii}{[N\,{\scriptsize II}]\,}
\theoremstyle{thmstyleone}%
\theoremstyle{thmstyletwo}%

\theoremstyle{thmstylethree}%

\begin{document}

\title[Article Title]{An extreme ram-pressure stripping event in a protocluster at redshift 4.3}

\author*[1]{\fnm{Dazhi} \sur{Zhou}}\email{dzhou.astro@gmail.com}
\author[2,3,1,4]{\fnm{Scott C.} \sur{Chapman}}
\author[5,6]{\fnm{Manuel} \sur{Aravena}}
\author[7,8,9]{\fnm{Roger} \sur{Deane}}
\author[10]{\fnm{Anthony H.} \sur{Gonzalez}}
\author[1]{\fnm{Ryley} \sur{Hill}}
\author[11]{\fnm{Nicholas} \sur{LeVar}}
\author[12]{\fnm{Matthew A.} \sur{Malkan}}
\author[13]{\fnm{Adam} \sur{Muzzin}}
\author[14]{\fnm{Nan} \sur{Zhang}}
\author[15,16,17]{\fnm{Kedar A.} \sur{Phadke}}
\author[1]{\fnm{Vismaya R.} \sur{Pillai}}
\author[18]{\fnm{Manuel} \sur{Solimano}}
\author[11]{\fnm{Justin S.} \sur{Spilker}}
\author[19]{\fnm{Nikolaus} \sur{Sulzenauer}}
\author[15,16,14]{\fnm{Joaquin D.} \sur{Vieira}}
\author[15]{\fnm{David} \sur{Vizgan}}
\author[1]{\fnm{George~C.~P.} \sur{Wang}}
\author[19]{\fnm{Axel} \sur{Weiss}}

\affil*[1]{\ubc}
\affil[2]{\orgdiv{Department of Physics and Atmospheric Science}, \orgname{Dalhousie University}, \country{Canada}}
\affil[3]{\orgdiv{National Research Council}, \orgname{Herzberg Astronomy and Astrophysics}, \country{Canada}}
\affil[4]{\orgname{Eureka Scientific Inc, Oakland, CA 94602, USA}}
\affil[5]{\orgdiv{Instituto de Estudios Astrof\'{\i}cos, Facultad de Ingenier\'{\i}a y Ciencias}, \orgname{Universidad Diego Portales, Av. Ej\'ercito 441, Santiago}, \country{Chile}}
\affil[6]{\orgdiv{Millenium Nucleus for Galaxies (MINGAL)}}
\affil[7]{\orgdiv{Inter-University Institute for Data Intensive Astronomy, \\Department of Astronomy}, \orgname{University of Cape Town, Rondebosch, Cape Town, 7701}, \country{South Africa}}
\affil[8]{\orgdiv{Wits Centre for Astrophysics, School of Physics}, \orgname{University of the Witwatersrand, 1 Jan Smuts Avenue, 2000, Johannesburg}, \country{South Africa}}
\affil[9]{\orgdiv{Department of Physics}, \orgname{University of Pretoria, Hatfield, Pretoria 0028}, \country{South Africa}}
\affil[10]{\ufastro}
\affil[11]{\tamu}
\affil[12]{\orgdiv{Department of Physics and Astronomy}, \orgname{University of California, 475 Portola Plaza, Los Angeles, CA 90095}, \country{USA}}
\affil[13]{\orgdiv{Department of Physics and Astronomy}, \orgname{York University, 4700 Keele St. Toronto, Ontario, M3J 1P3}, \country{Canada}}
\affil[14]{\illinoisphysics}
\affil[15]{\illinoisastro}
\affil[16]{\illinoiscaps}
\affil[17]{NSF-Simons AI Institute for the Sky (SkAI), 172 E. Chestnut St., Chicago, IL 60611, USA}
\affil[18]{\orgdiv{Centro de Astrobiologia (CAB)}, \orgname{CSIC-INTA, Ctra. de Ajalvir km 4, Torrejon de Ardoz, E-28850, Madrid}, \country{Spain}}
\affil[19]{\mpifr}

\newcommand{\illinoisastro}{\text{Department of Astronomy, University of Illinois, 1002 West} \linebreak \text{Green St., Urbana, IL 61801, USA}}
\newcommand{\illinoiscaps}{\orgdiv{Center for AstroPhysical Surveys}, \orgname{National Center for Supercomputing Applications, 1205 West Clark Street, Urbana, IL 61801}, \country{USA}}
\newcommand{\illinoisphysics}{\orgdiv{Department of Physics}, \orgname{University of Illinois, 1110 West Green St., Urbana, IL 61801}, \country{USA}}
\newcommand{\ufastro}{\text{Department of Astronomy, University of Florida, Gainesville, FL 32611, USA}}
\newcommand{\tamu}{\text{Department of Physics and Astronomy and George P. and Cynthia} \linebreak \text{Woods Mitchell Institute for Fundamental Physics and Astronomy, }\linebreak \text{Texas A\&M University, 4242 TAMU, College Station, TX 77843-4242, USA}}
\newcommand{\ubc}{\orgdiv{Department of Physics and Astronomy}, \orgname{University of British Columbia, 6225 Agricultural Rd., Vancouver, V6T 1Z1}, \country{Canada}}
\newcommand{\mpifr}{\text{Max-Planck-Institut für Radioastronomie, Auf dem Hügel 69,}\linebreak \text{53121, Bonn, Germany}}
\maketitle
\textbf{In the nearby Universe, the environment plays a crucial role in suppressing star formation in dense regions. 
In particular, ram-pressure stripping (RPS) is a major mechanism for removing gas from galaxies in clusters, occurring when galaxies travel through a dense hot atmosphere and leave trailing gaseous wakes. 
By depleting the cold gas reservoir, RPS can drive outside-in quenching and is therefore thought to be an important route for transforming cluster galaxies. 
At earlier times, however, the hot atmosphere in protoclusters is expected to be immature, so environmental effects are commonly assumed to be dominated by gravitational interactions. 
Here we report ALMA and {\it JWST} observations of SPT2349$-$56-C26 (hereafter C26), a massive galaxy experiencing an extreme RPS event in the SPT2349$-$56 protocluster at $z\,{=}\,4.3$. 
More than half of the \cii-traced cold gas lies outside its stellar body, with the emission peak offset by 6\,kpc. 
These observations show that RPS can remove most of the cold gas from massive galaxies in dense protocluster cores as early as $z=4.3$, providing a direct hydrodynamic pathway for environmental quenching at $z>4$.  }
\smallskip

SPT2349$-$56 is a massive protocluster containing about 30 dusty star-forming galaxies (DSFGs) within a 100-kpc region \cite{Vieira2010,Miller2018,Hill2020,Sulzenauer2025,Chapman2025}. 
The recent detection of the thermal Sunyaev-Zeldovich (SZ) effect with the Atacama Large Millimeter/submillimeter Array (ALMA) suggests that intracluster gas is already hot and abundant in this system \cite{Zhou2026}. 
This makes the SPT2349$-$56 protocluster a rare laboratory for studying environmental processing before galaxy clusters become fully relaxed \cite{Fujita2004,Sarron2019}. 

\section*{Results}
C26 was originally identified in ALMA observations of the \cii158\,$\mu$m line as an extended gas structure in the SPT2349$-$56 protocluster \cite{Sulzenauer2025}, which lies at a projected distance of about 78\,kpc ($\approx R_{500}$) from the tSZ centroid. 
Follow-up James Webb Space Telescope ({\it JWST}) imaging then revealed the characteristic cometary stellar morphology of the system, consisting of a compact head and an elongated tail (Fig.\,\ref{fig1}) \cite{Kapferer2009}. 
The \cii\ emission peak is offset by 6\,kpc from the stellar head seen in the {\it JWST} images, revealing a strong decoupling between the gas and stellar morphologies (Fig.\,\ref{fig2}a). 
In the high-resolution NIRCam F200W image, the tail is clumpy and extends about 20\,kpc from the head, with a position angle aligned with the cluster-centric axis. 
A bright downstream knot is visible in the lower-resolution NIRCam F444W and MIRI F1000W images at a projected distance of $\sim$\,12\,kpc (Fig.\,\ref{fig1}). 

\subsection*{Stellar properties}
We investigate the stellar properties of the head, tail, and downstream knot using 6-band photometry from Hubble Space Telescope ({\it HST}) and {\it JWST} (Extended Data Fig.\,\ref{exfig1}; Extended Data Table \ref{extable2}). 
Relative to the head, the tail is bluer and clumpier, and is detected in the rest-frame ultraviolet (UV) bands (Extended Data Fig.\,\ref{exfig2} and \ref{exfig3}), suggesting a younger stellar population. 
SED fitting indicates that C26 consists of a massive head 
($M_\star = 2.2^{+0.7}_{-0.5}\times10^{10}\,{\rm M_\odot}$) 
and a lower-mass tail 
($M_\star = 5.9^{+1.6}_{-1.3}\times10^{9}\,{\rm M_\odot}$), which includes the downstream knot
($M_\star = 3.2^{+0.8}_{-0.8}\times10^{9}\,{\rm M_\odot}$; Extended Data Table \ref{extable3}). 
Compared with typical main-sequence galaxies at $z\sim4.5$ \cite{Faisst2020, bethermin2020}, the stellar head has a $\sim 0.5$ dex lower specific star-formation rate (${\rm sSFR}=1.33^{+0.55}_{-0.49}\,{\rm Gyr^{-1}}$), while the values of the tail ($2.99^{+2.97}_{-1.10}\,{\rm Gyr^{-1}}$) and knot ($2.37^{+1.85}_{-1.04}\,{\rm Gyr^{-1}}$) are slightly lower but broadly consistent within the scatter. 

\subsection*{Cold-gas reservoir and excitation}
We quantify the cold-gas reservoir using \cii, dust continuum, and CO emission as cold-gas tracers (Methods; Extended Data Fig.\,\ref{exfig4}; Extended Data Table \ref{extable1} and \ref{extable4}). 
The \cii\ emission is detected across the whole system ($L_{\rm \cii}^{\rm full}=(1.17\pm0.17)\times10^9\rm\,L_\odot$, $L_{\rm \cii}^{\rm head}=(4.8\pm0.7)\times10^8\rm\,L_\odot$, $L_{\rm \cii}^{\rm tail}=(6.5\pm0.9)\times10^8\rm\,L_\odot$), with the tail containing at least $55\%$ of the total \cii\ emission. 
Adopting $\alpha_{\rm [CII]}=30\,{\rm M_\odot\,L_\odot^{-1}}$ \cite{Zanella2018}, this corresponds to $M_{\rm mol}^{\rm full}=(3.5\pm0.5)\times10^{10}\rm\,M_\odot$ for the full system, $M_{\rm mol}^{\rm head}=(1.4\pm0.2)\times10^{10}\rm\,M_\odot$ for the head, and $M_{\rm mol}^{\rm tail}=(2.0\pm0.3)\times10^{10}\rm\,M_\odot$ for the tail. 
The detection of the 850-$\mu$m continuum emission yields a slightly lower dust-based gas mass of $(2.4\pm0.5)\times10^{10}\,{\rm M_\odot}$ \cite{Scoville2016}. 
The CO(2--1) detection gives $L'_{\rm CO(2-1)}=(3.0\pm1.0)\times10^{10}\rm\,K\,km\,s^{-1}pc^2$, corresponding to $M_{\rm mol}^{\rm CO(2-1)}=(2.8\pm0.9)\times10^{10}\,{\rm M_\odot}$ for $\alpha_{\rm CO}=0.8\,{\rm M_\odot\,(K\,km\,s^{-1}\,pc^2)^{-1}}$ and $r_{21}=0.85$ \cite{Carilli2013,Aravena2016}, which is consistent with the \cii\ and dust continuum results within uncertainties. 
In contrast, no CO(4--3) emission is detected in the deep ALMA data, implying an unusually low excitation ratio $r_{42}<0.13$ (Methods). 
The gas is therefore likely in low excitation with low star-formation activity, rather than dominated by warm molecular clouds \cite{ivison2011,Carilli2013}. 

\section*{Discussion}
The combined evidence strongly favors RPS as the dominant mechanism responsible for the displaced cold gas in C26. 
The \cii\ emission peak is offset by 6\,kpc from the stellar head centroid, with more than half of the \cii-traced cold gas lying outside the main stellar body, suggesting the strong decoupling between the gaseous and stellar components. 
Moreover, the smaller displacement of the dust continuum (4\,kpc; Fig.\,\ref{fig2}c) further suggests phase-dependent stripping, in which the more diffuse gas is stripped more efficiently than the denser dust-traced material \cite{Boselli2022}. 
The tail is also aligned with the cluster-centric direction \cite{Salinas2024}, while the independently detected hot ICM in the protocluster provides the external medium required for ram pressure \cite{Zhou2026}. 

The ionized gas provides additional evidence that the displaced material is interacting with the surrounding medium (Extended Data Fig.\,\ref{exfig5}). 
The \nii205\,$\mu$m emission is enhanced on both flanks of the tail (Fig.\,\ref{fig2}d). 
If attributed entirely to H\,II-region emission, its luminosity would imply an SFR of $\sim223\,M_\odot\,{\rm yr}^{-1}$ using the local \nii--\,SFR relation \cite{Zhao2016}, far above the SED-inferred value for the tail ($18.3^{+13.9}_{-6.9}\,M_\odot\,{\rm yr}^{-1}$). 
Star formation alone is therefore unlikely to account for all of the \nii\ emission, while its flank-brightened morphology is consistent with ionized interface layers around the displaced cold gas caused by its interaction with the ambient hot gas \cite{Poggianti2019,Sun2021}. 

Although gravitational interactions cannot be excluded completely, they are less likely to dominate the observed structure according to the available evidence. 
A major merger is disfavored because the downstream knot contains less than 15\% of the stellar mass of the head, making it unlikely to drive the observed large-scale displacement of such a massive gas reservoir. 
The \cii\ emission remains kinematically connected from the head through the tail without a clearly detached component (Fig.\,\ref{fig2}b and Extended Data Fig.\,\ref{exfig6}), indicating that the knot and tail belong to the same gaseous structure. 
The CO observations further suggest that the molecular gas is predominantly in a very low-excitation state, broadly consistent with the diffuse, sub-thermally excited gas seen in the tails of low-redshift jellyfish galaxies \cite{Moretti2018,Moretti2020}. 
Strong gravitational interactions commonly produce enhanced central star formation and elevated molecular-gas excitation, whereas C26 shows only modest star formation and no CO(4--3) detection in either the head or the tail, despite still containing a substantial cold-gas reservoir. 
For the gas mass of the stellar head, a typical high-redshift excitation condition would predict CO(4--3) emission detectable at $\gtrsim8\sigma$ in our deep ALMA Band-3 data. 
Its non-detection therefore argues against a strongly interaction-driven state of the main galaxy. 
While a past tidal encounter or high-velocity flyby cannot be entirely ruled out, the current observations collectively favor hydrodynamic stripping as the dominant mechanism shaping C26. 

The large displacement of the cold-gas reservoir in C26 stands out when compared with ram-pressure stripped galaxies previously observed at low redshift \cite{Moretti2018,Jachym2019,Ramatsoku2020}. 
Despite being a massive, gas-rich system, the \cii\ emission peak shows a 6-kpc offset from the stellar head and more than half of the \cii\ emission lies outside the main stellar body. 
In local jellyfish galaxies, the stripped gas tail can extend far beyond the stellar disk, but the bulk of the gas emission usually remains associated with the main galaxy. 
The typical centroid offsets between ionized-gas emission and stellar continuum are about $\sim$\,1\,kpc \cite{Liu2021}, which is more modest in denser cold-gas tracers with their emission peaks within the stellar disk \cite{Moretti2018,Jachym2019,Moretti2020,Ramatsoku2020, Deb2020}. 
The large separation in C26 suggests unusually severe cold-gas removal, hinting at stronger dynamical stripping than is commonly inferred in nearby jellyfish systems. 

The extreme stripping seen in C26 shows that external gas removal can be even more vigorous before the clusters reach their more mature state at $z\,{\approx}\,2$.
This is in contrast to the common expectation that ICM-driven environmental effects are weaker in high-redshift protoclusters, before the establishment of a mature ICM \cite{Dannerbauer2017,Wang2018b,Momose2024,pm2025}. 
Such unusually strong ram pressure is physically plausible in SPT2349$-$56 because the hot ICM at $z=4.3$ is expected to be substantially denser than in comparable lower-redshift systems (Methods, Fig.\,\ref{fig4}). 
Given that the system is dynamically active and not yet fully relaxed \cite{Hill2020}, ICM substructure and cluster assembly could further enhance RPS through local ICM compression and increased relative velocity between galaxies and the ambient gas \cite{Tonnesen2008, Ebeling2019, Ruggiero2019}. 
The presence of an over-heated ICM, together with enhanced radio-AGN activity in the protocluster core \cite{Zhou2026,Chapman2025}, suggests that the surrounding medium may be dynamically disturbed, with possible turbulence, shocks, winds, or bulk motions \cite{saha2026}. 
A denser and possibly more turbulent hot ICM in such a dynamically assembling system would make RPS more effective for infalling galaxies \cite{Li2018, Merluzzi2024, Ge2026}. 

C26 demonstrates that, at $z>4$, RPS can remove most of the cold-gas reservoir from a massive galaxy while the stellar body is still forming stars. 
Recent studies suggest that RPS can already occur before mature cluster assembly \cite{Xu2025,Li2026,Doven2026}. 
In the forming cluster CLJ1001 at $z=2.51$, RPS galaxies mostly show enhanced star-formation activity and higher star-formation efficiency, consistent with compression-triggered star formation in the disturbed gas \cite{Xu2025}. 
In contrast, the post-starburst galaxy A2744-JF-z3 at $z=3.06$ shows evidence of RPS together with an abrupt cessation of star formation \cite{Li2026}. 
C26 may capture an intermediate stage between compression-enhanced star formation and complete quenching, in which most of the cold-gas reservoir has already been removed by the external environment, while the stellar head is not yet fully quenched. 
More than half of the observed cold gas is already displaced from the galaxy body, and the reduced gas mass correspondingly lowers the gravitational restoring force of the remaining head. 
Ram-pressure stripping may therefore continue to remove the remaining cold gas, cutting off the fuel supply required to sustain star formation in the stellar head. 
C26 therefore provides direct evidence that ram-pressure stripping can dominate the quenching of a massive star-forming galaxy by removing its cold gas reservoir, linking early stripping to the emergence of high-redshift quiescent galaxies in dense environments (Fig.~\ref{fig5}). 

If C26 is not unique, this mechanism may already be transforming other galaxies in the SPT2349$-$56 core. 
Compared to field galaxies, protocluster galaxies in SPT2349$-$56 have on average lower gas fractions and higher gas excitation temperatures \cite{Hill2020,Hill2022,Hughes2024}. 
About 40\% of the cold-gas emission is not associated with individual galaxies, but instead resides in a diffuse component, with giant gas streamers present in the protocluster core \cite{Zhou2025,Harrington2025,Sulzenauer2025}. 
Moreover, a strong concentration of red compact galaxies has recently been identified in the core without \cii\ or dust-continuum detections, suggesting that a substantial gas-poor population is already present in SPT2349$-$56 \cite{chapman2026}. 
Ram-pressure stripping may therefore already be contributing to the formation of this gas-poor population, while redistributing baryons into the intracluster environment. 
Environmental stripping thus provides a direct pathway for the early suppression of star formation while redistributing chemically enriched baryons into the intracluster environment and potentially contributing to the early enrichment of the nascent ICM \cite{Domainko2006,Chen2024,Tanaka2024,Zhou2025}. 
High-resolution ALMA and {\it JWST} follow-up observations will further clarify the interplay between the hot medium and the displaced gas, providing stronger constraints on environmental processes in young galaxy clusters.
 
\clearpage

\begin{figure}[!h]
    \centering
    \includegraphics[width=0.98\linewidth]{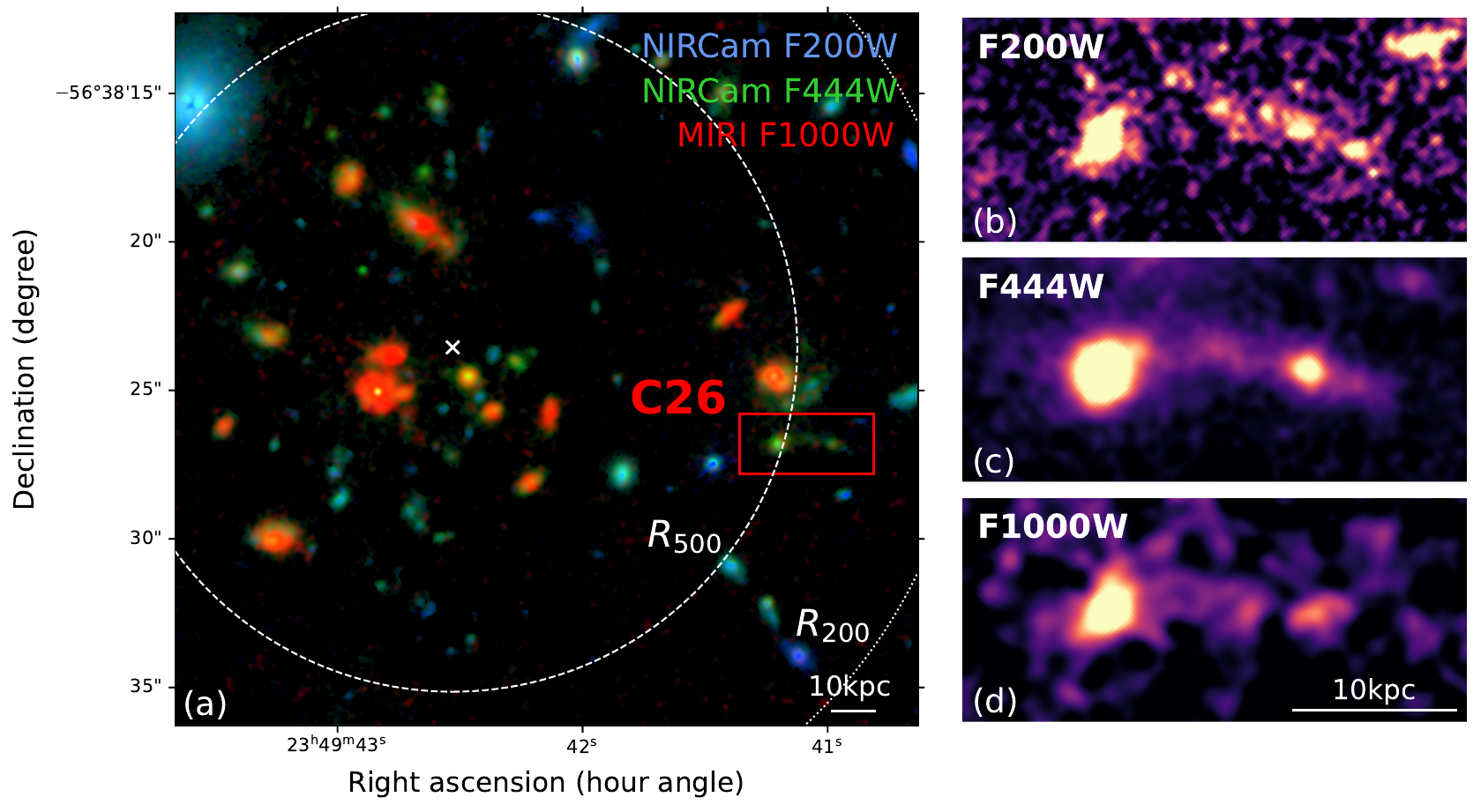}
    \caption{{\bf {\it JWST} imaging of the RPS galaxy C26 in SPT2349$-$56.} \\
    (a) Red-green-blue image of the SPT2349$-$56 protocluster core (blue: NIRCam/F200W; green: NIRCam/F444W; red: MIRI/F1000W). 
    The RPS galaxy C26 is marked by the red rectangle. 
    The cross labels the tSZ centroid of the protocluster. 
    $R_{200}$ and $R_{500}$ are indicated in dotted and dashed lines. 
    (b--d) Zoomed images of C26 in F200W, F444W, and F1000W, respectively. 
    Scale bars show 10 kpc at $z\,{=}\,4.3$. 
    }
    \label{fig1}
\end{figure}

\begin{figure}[htp]
    \centering
    \includegraphics[width=\linewidth]{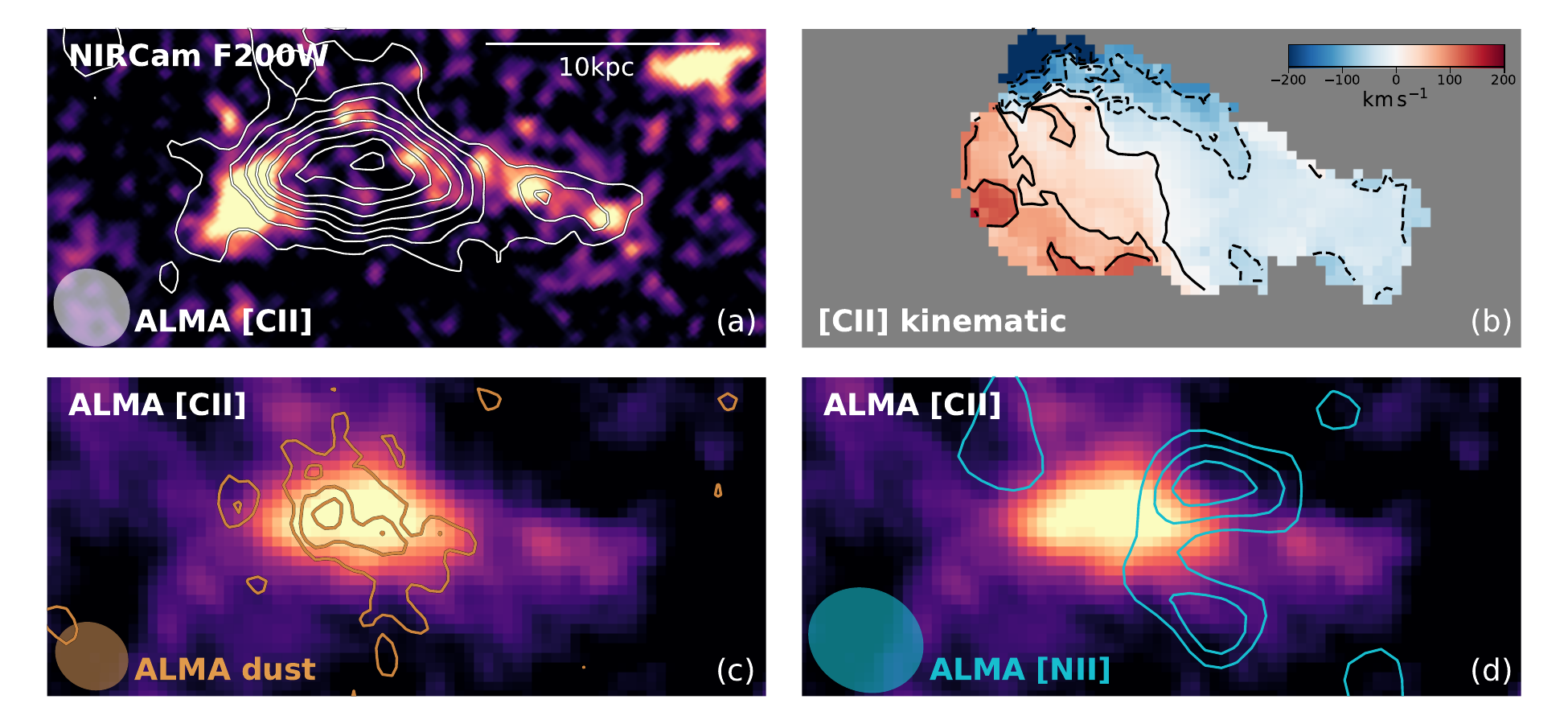}
    \caption{{\bf ALMA imaging of the multiphase ISM in C26.}\\
    (a) NIRCam F200W image with ALMA \cii\ moment-0 contours at [3, 4.5, 6, 7.5, 9, 10.5, 12, 13.5] times the rms level. 
    (b) ALMA \cii\ moment-1 map with contours at [-200, -150, -100, -50, 0, 50, 100, 150, 200]\,km\,s$^{-1}$. 
    The noisy pixels are masked based on their corresponding channel significance (Methods). 
    (c) ALMA 850-$\mu$m dust continuum contours overlaid on the ALMA \cii\ moment-0 map. 
    The brown contours are [2,3,4] times the rms level of the ALMA dust continuum map. 
    (d) ALMA \nii\ 205\,$\mu$m moment-0 contours overlaid on the ALMA \cii\ moment-0 map. 
    The cyan contours are [2,3,4] times the rms level of the ALMA \nii\ moment-0 map. }
    \label{fig2}
\end{figure}

\clearpage

\begin{figure}[htp]
    \centering
    \includegraphics[width=0.98\linewidth]{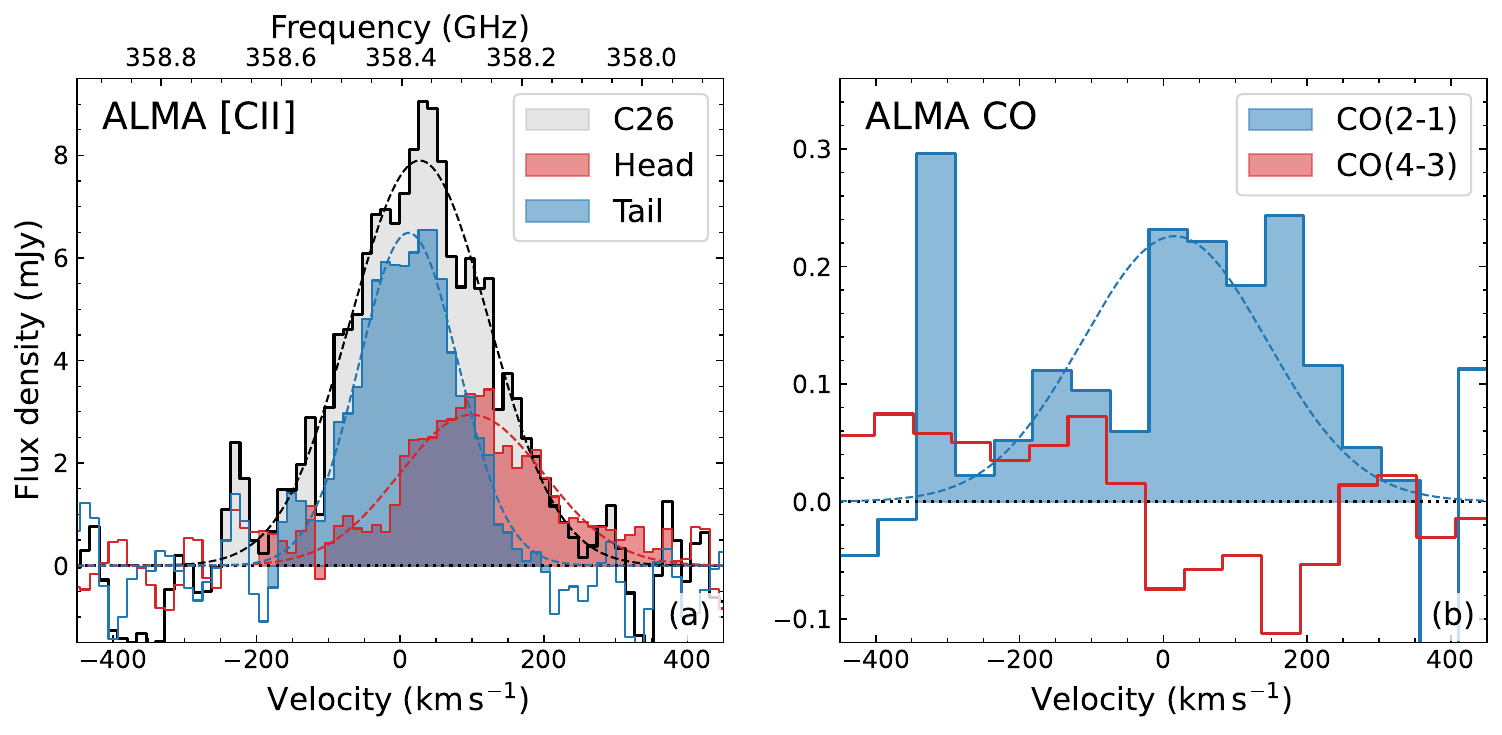}
    \caption{{\bf ALMA \cii\ and CO spectra of C26. } \\
    (a) ALMA \cii\ spectra extracted for the full system (grey), the stellar head (red), and the tail (blue). 
    (b) ALMA CO(2--1) (blue) and CO(4--3) (red) spectra measured over the same spatial aperture for the full C26 system. 
    The dashed curves are the best-fit Gaussian profiles for determining the velocity ranges for velocity-integrated intensity calculations, which are indicated as the shaded regions. 
    }
    \label{fig3}
\end{figure}

\clearpage
\begin{figure}
    \centering
    \includegraphics[width=0.98\linewidth]{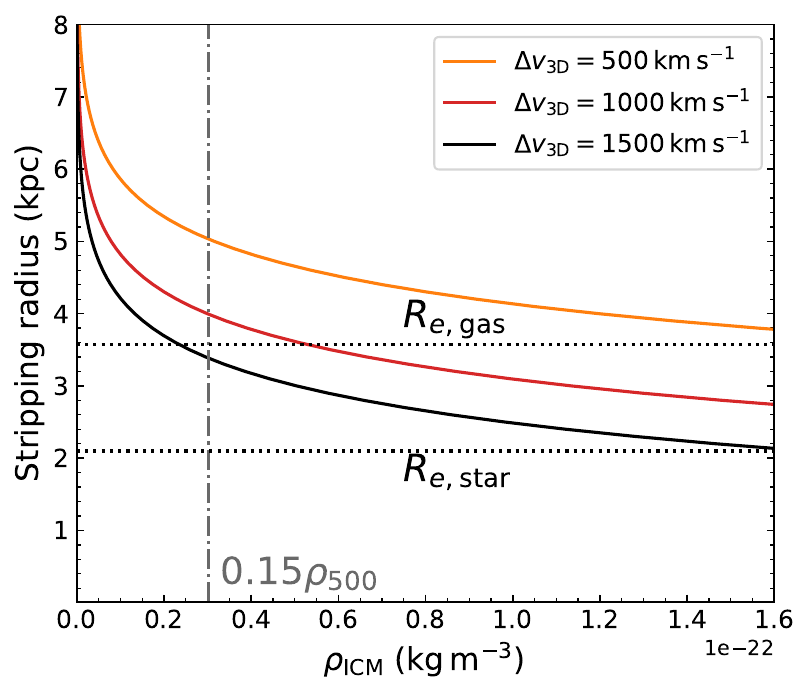}
    \caption{{\bf Initial stripping radius as a function of ambient ICM density.}\\
    Solid curves show the initial stripping radius estimates for three assumed three-dimensional differential velocities related to the ICM. 
    The vertical dot-dashed line indicates a gas density corresponding to 15\% of the characteristic overdensity ($\rho_{500}$) at $z\,{=}\,4.3$. 
    Horizontal dotted lines indicate the half-light radius of the stellar head measured from the F444W image and the expected half-light radius of the gaseous head before stripping, extrapolated from the empirical relation \cite{Boselli2006}. 
    }
    \label{fig4}
\end{figure} 
\clearpage

\begin{figure}[!h]
    \centering
    \includegraphics[width=0.98\linewidth]{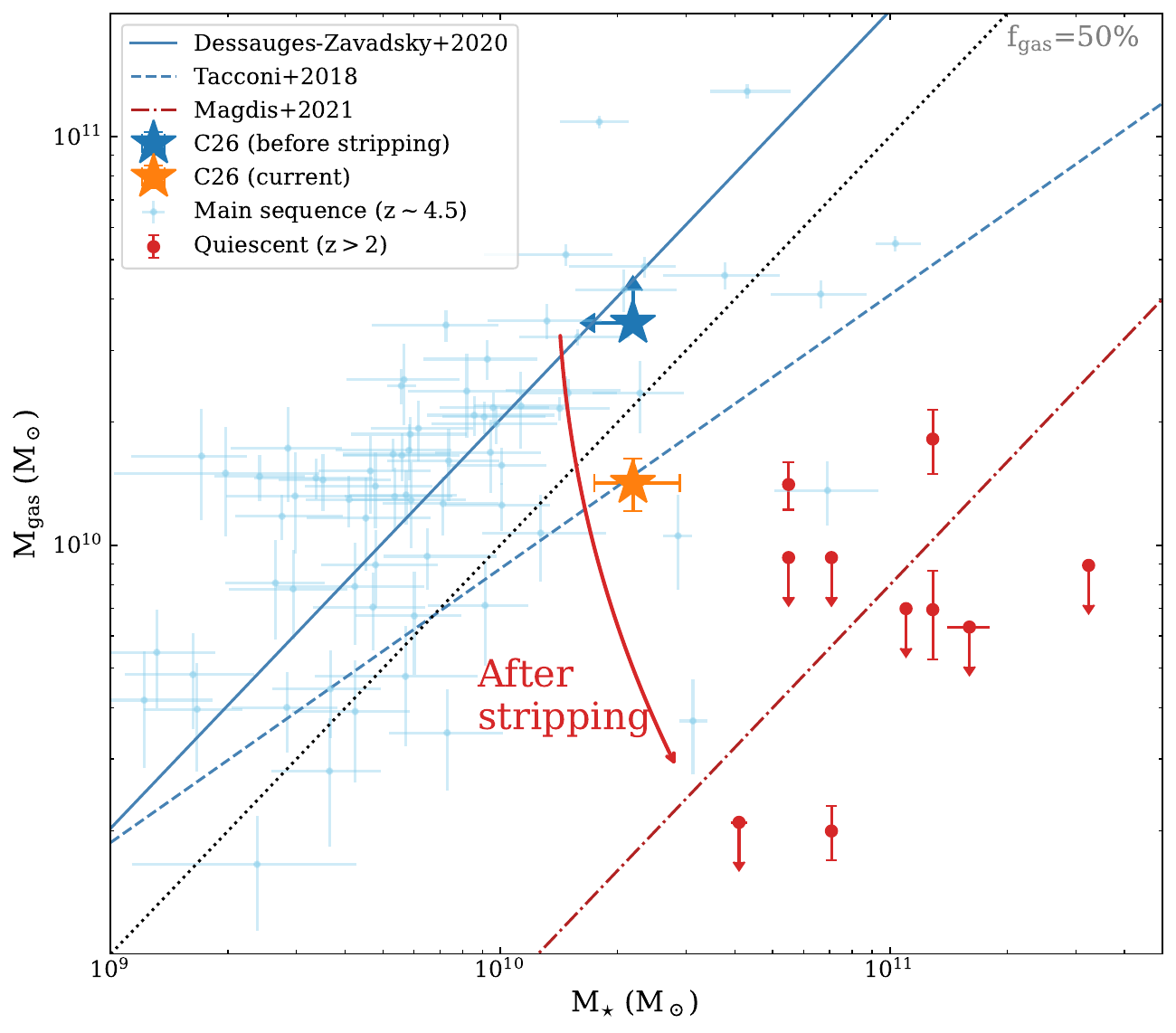}
    \caption{{\bf Ram-pressure stripping drives C26 toward a gas-poor state} \\
    C26 in the $M_{\rm gas}-M_\star$ plane, compared with main-sequence galaxies at $z\sim4.5$ (blue) and quiescent galaxies at $z>2$ (red) \cite{bethermin2020, Faisst2020, Wang2025, Umehata2025, Scholtz2026,Adscheid2025,DEugenio2026}. 
    The blue star represents the inferred pre-stripping state of C26, using the cold-gas mass of the full system and the stellar mass of the head. 
    The orange star marks the current state of the stellar head with its remaining cold gas during RPS.
    Literature relations are shown in blue for star-forming galaxies and in red for quiescent galaxies \cite{Dessauges-Zavadsky2020, Tacconi2018, Magdis2021}. 
    The dotted line shows a gas fraction of 50\%. 
    Error bars for C26 show the 16th-84th percentile uncertainty in $M_\star$ and the $1\sigma$ measurement uncertainty in the \cii-based $M_{\rm gas}$. 
    Uncertainties for the literature samples are shown as reported in the corresponding studies, with downward arrows indicating upper limits.
    The arrow schematically illustrates a possible future evolutionary path for C26, in which continued hydrodynamic stripping progressively depletes the cold-gas reservoir while leaving the stellar component largely unaffected, eventually promoting the transition from a gas-rich phase toward a quiescent state.
    }
    \label{fig5}
\end{figure}
 
\clearpage
\section*{Methods}\label{sec11}

Throughout this paper, we use the standard Lambda Cold Dark Matter cosmological model with $H_0\,{=}\,67.7\,\rm km\,s^{-1}\,Mpc^{-1}$ and $\Omega_m\,{=}\,0.31$ \cite{planck2015}, which corresponds to a proper angular scale of 6.9\,kpc\,arcsec$^{-1}$ at $z\,{=}\,4.3$. 

\subsection*{Observations and data reduction} \label{obs}

\subsubsection*{{\it HST} and {\it JWST} imaging}

Hubble Space Telescope ({\it HST}) WFC3/IR F110W and F160W (PID: 15701, PI: S.\,Chapman) and James Webb Space Telescope ({\it JWST}) NIRCam F200W, F444W, and MIRI F1000W, F1800W (PID: 06669, PI: S.\,Chapman) images of the SPT2349$-$56 protocluster were obtained from the Mikulski Archive for Space Telescopes (MAST) server (\hyperlink{https://mast.stsci.edu/portal/Mashup/Clients/Mast/Portal.html}{https://mast.stsci.edu}). 
The exposure times are 5674\,s (F110W), 8472\,s (F160W), 3264\,s (F200W), 3264\,s (F444W), 1454\,s (F1000W), and 4029\,s (F1800W). 

We followed the standard reduction procedure to process the level-1 data through the {\it HST} and {\it JWST} calibration pipelines ({\tt hstcal/calwf3} and {\tt jwst/calwebb} \cite{jwstpipe}) within the Space Telescope Environment ({\tt stenv}). 
We applied an extra de-striping step for the {\it JWST} images during the stage 2 process to mitigate the influence of column stripes and reduce the $1/f$ noise. 
For the F200W image, we independently estimated the background in each region defined by the weight map to correct the inhomogeneous sky background across the detector gap due to low coverage. 
We first aligned the F444W image to the Gaia DR3 catalog in stage 3 during the {\tt tweakreg} step \cite{gaiadr3}. 
Other bands were then aligned to F444W using the python packages {\tt tweakreg} in {\tt general} mode. 
The astrometry uncertainty is 0.02\,arcsec--0.04\,arcsec for each band. 

The reduced images reach 5$\sigma$ depths down to F110W$\,{=}\,$28.0\,mag, F160W$\,{=}\,$27.6\,mag, F200W$\,{=}\,$27.8\,mag, F444W$\,{=}\,$28.3\,mag, F1000W$\,{=}\,$26.4\,mag, and F1800W$\,{=}\,$25.7\,mag (AB magnitudes), which were measured based on the standard deviations of fluxes within a 0.4\,arcsec circular aperture placed at 10000 random positions in the 2$\sigma$ clipped image of each band.

Next, we reprojected all images to a common grid defined by the F200W image using the {\tt reproject\_adaptive} function from the {\tt reproject} package, with {\tt conserve\_flux} enabled for the {\it HST} images \cite{reproject}. 
The images have a pixel size of 0.0312\,arcsec and are shown in Extended Data Fig.\,\ref{exfig1}. 
The red-green-blue image was then produced from the F200W, F444W, and F1000W images with the {\tt astropy} function {\tt make\_lupton\_rgb}, which is presented in Fig.~\ref{fig1}. 
For demonstration purposes, we mildly smoothed the F200W and F1000W images by 2D Gaussian kernels with kernel sizes of 0.7 and 1 pixels, respectively.

\subsubsection*{ALMA observations}

In this study, we utilized ALMA observations of \cii 158\,$\mu$m, \nii 205\,$\mu$m, CO(2--1), CO(4--3), and the 850-$\mu$m and 3.2-mm continuum from the ALMA Science Archive to investigate the multiphase ISM associated with C26 (Extended Data Table \ref{extable1}). 
We used the standard ALMA calibration script to uniformly recalibrate the obtained ALMA data with the latest {\tt CASA} \cite{CASA} pipeline ({\tt CASA} version 6.6.6). 
For the luminous DSFGs in SPT2349$-$56, the imaging sensitivity is limited by dynamical range instead of thermal noise in Band 7. 
We therefore performed phase-only self calibration for Band 7 data using the continuum emission of protocluster DSFGs. 
The suppressed phase variations reduce dirty sidelobe residuals from bright sources, allowing deeper deconvolution to recover faint emission. 

We used a two-step cleaning strategy to produce the \cii, \nii, CO(2--1), and CO(4--3) data cubes in {\tt CASA}. 
First, we used the {\tt CASA} task {\tt tclean} to deconvolve the data cubes to {\tt nsigma=4} with the {\tt multiscale} deconvolver ({\tt smallscalebias=0.8}, {\tt scales=[0, 5, 10]}) in a common synthesized beam size. 
We chose the {\tt mosaic} gridder, {\tt mosweight=True}, Briggs weighting ({\tt robust=0.5} for \cii, \nii, and CO(2--1) and {\tt robust=-0.5} for CO(4--3) imaging) without a cleaning mask for this shallow deconvolution. 
The pixel sizes are 0.06, 0.10, 0.17, and 0.20\,arcsec for \cii, \nii, CO(2--1), and CO(4--3) respectively. 
These shallowly cleaned maps were used to construct a high-confidence clean mask similar to the CPROPS mask \cite{Rosolowsky2006,Leroy2021}. 
We selected all regions above 5 times the rms or ${>}\,4\sigma$ with an adjacent channel ${>}\,3\sigma$ to create the initial mask. 
The initial mask was then dilated by one synthesized beam and by one channel to include neighborhood pixels with ${>}\,2\sigma$. 
To capture the faint wings of the emission lines, we further grew the mask spectrally by 2 channels. 
We used this mask to clean the data cube deeper down to {\tt nsigma=1} with the {\tt hogbom} deconvolver. 
This cleaning strategy yields a central rms of 0.13\,mJy\,beam$^{-1}$, 0.13\,mJy\,beam$^{-1}$, 77\,$\mu$Jy\,beam$^{-1}$, and 58\,$\mu$Jy\,beam$^{-1}$ for a channel width of 13\,km\,s$^{-1}$, 17\,km\,s$^{-1}$, 54\,km\,s$^{-1}$, and 54\,km\,s$^{-1}$ for the \cii, \nii, CO(2--1), and CO(4--3) data, respectively. 
Their corresponding synthesized beam sizes are 0.51\,arcsec$\,{\times}\,$0.45\,arcsec, 0.74\,arcsec$\,{\times}\,$0.64\,arcsec, 0.98\,arcsec$\,{\times}\,$0.85\,arcsec, and 1.30\,arcsec$\,{\times}\,$1.12\,arcsec. 
While the CO(2--1) and CO(4--3) data have sufficient line-free channels to estimate the continuum level for continuum subtraction in the $uv$ plane, the same strategy is not feasible for the \cii\ and \nii\ data cubes due to the limited line-free channels caused by the wide velocity distribution of the protocluster galaxies. 
We therefore obtained the \cii\ and \nii\ continuum maps from the median values along the spectral axis from the $2\sigma$ clipped data cubes and subsequently subtracted the continuum from their corresponding data cubes. 

A similar two-step cleaning strategy was used to image the 850-$\mu$m continuum map. 
We first flagged channels with strong \cii\ emission from 357.3\,GHz to 359.8\,GHz. 
The remaining channels were imaged with the {\tt CASA} task {\tt tclean}. 
We used the {\tt mosaic} gridder, {\tt mosweight=True}, Briggs weighting ({\tt robust=0.5}), and {\tt multiscale} deconvolver with no mask and cleaned down to {\tt nsigma=4}. 
We built the initial mask from pixels with signals above 4$\sigma$ and then expanded it by one synthesized beam to include adjacent pixels with values ${>}\,2\sigma$. 
Then, we performed single-scale cleaning with the {\tt hogbom} deconvolver to further clean down to {\tt nsigma=0.5} within the mask. 
The final 850-$\mu$m continuum map reaches an rms of 11\,$\mu$Jy\,beam$^{-1}$ at the phase center with a synthesized beam size of 0.47\,arcsec$\,{\times}\,$0.41\,arcsec. 

The 3.2-mm continuum image was obtained in a similar manner, largely following the procedure described in our previous work \cite{Zhou2026}. 
We utilized all possible ALMA Band-3 data and used line-free channels for the continuum imaging with Briggs weighting ({\tt robust=0.5}). 
The visibilities with $uv$ distance ${<}\,30\,k\lambda$ were excluded for a smaller synthesized beam size and to remove the contamination from the large-scale SZ decrement. 
The 3.2-mm continuum map reaches an rms of 2.6\,$\mu$Jy\,beam$^{-1}$ with a synthesized beam of 1.13\,arcsec$\,{\times}\,$1.16\,arcsec. 

\subsection*{Photometry and SED modelling}\label{pho}

\subsubsection*{Component photometry} 

We used the {\tt photutils} package \cite{photutils} for photometry measurements. 
We subtracted the sky background using the {\tt Background2D} function with the {\tt MedianBackground} estimator. 
For each band, we visually identified good stars from the catalog obtained by the {\tt find\_peaks} algorithm and used {\tt EPSFBuilder} to construct empirical PSF models \cite{Anderson2000}. 
We convolved the data with the kernel obtained from the {\tt create\_matching\_kernel} function with a low-pass window function {\tt SplitCosineBellWindow} to match the point spread functions (PSFs) to the F1000W PSF for all images except F1800W, for which the PSF is too sparse. 
The effective FWHM of the matched PSF is 0.398\,arcsec. 

We used the F444W band as our detection image. 
The threshold image was defined as 1.5 times the F444W rms from {\tt Background2D}. 
We constructed a segmentation map using the {\tt detect\_sources} function for sources spanning more than 10 connected pixels above the detection threshold ({\tt n\_pixels=10}), and used {\tt deblend\_sources} ({\tt contrast=0}, {\tt nlevels=100}) to separate the head and tail. 
The deblending algorithm separated the tail into two segments, which we manually reassigned to a single segmentation label for subsequent photometric measurements. 
We then ran {\tt SourceFinder} and derived the corresponding Kron apertures with parameters {\tt Rmin=1.7} and {\tt k=1.2}, as shown in Extended Data Fig.\ref{exfig2}. 
These apertures were used to measure the fluxes of the head and tail in each band, with nearby sources masked to prevent contamination. 
Because the downstream knot is embedded within the tail and could not be cleanly isolated by the segmentation-based procedure, we instead measured its fluxes     using fixed circular aperture photometry with a radius of $0.3''$ centered on the F444W knot defined in the F444W image. 
The photometry uncertainties were estimated from the flux distribution of 10,000 randomly placed apertures on the 2$\sigma$-clipped image. 
Flux losses outside the apertures were corrected assuming a point-source profile based on our empirical matched PSF.
For the F1800W band, we independently corrected the aperture loss using its own empirical PSF. 
The final photometric catalog for the head, tail, and knot is given in Extended Data Table\,\ref{extable2}. 

\subsubsection*{SED fitting}\label{sed}

We derived the physical properties of the head, tail and knot with the {\tt MAGPHYS} spectral energy distribution (SED) fitting code \cite{dacunha2008}. 
We used the {\tt MAGPHYS} high-$z$ extension (v2), which accounts for the strong ultraviolet absorption feature at 2175\,\AA \cite{dacunha2015,Battisti2020}. 
The redshift was fixed to $z\,{=}\,4.303$ determined by the \cii\ line. 
Given the spatial offset between dust and stellar emission, we excluded the far-infrared measurements and only included the 6-band photometric measurements from {\it HST} and {\it JWST}. 
Additional 10\% uncertainties were added to account for the measurement error. 

The 16th, 50th, and 84th percentile values of the stellar mass, $A_{\rm v}$, star-formation rate (SFR), and specific SFR are listed in Extended Data Table \ref{extable3}. 
Both the tail and the embedded knot are bluer than the head, consistent with younger stellar populations formed in the stellar tail. 
Given the limited broad-band photometric coverage and the relatively restricted star-formation histories implemented in {\tt MAGPHYS}, the stellar masses of these young components remain uncertain and may be overestimated because recent star formation can dominate the stellar light and bias stellar-mass estimates \cite{Jones2026}. 
Even so, the downstream knot remains much less massive than the head, with a stellar mass ratio of ${<}\,0.15$. 

\subsection*{Cold-gas reservoir and ISM diagnostics}

We used the \cii, \nii, CO(2--1), CO(4--3), and 850-$\mu$m continuum data to examine the spatial distributions and luminosities of the neutral gas, ionized gas, molecular gas, and thermal dust emission within C26.

The \cii\ spectrum of C26 was extracted within the Kron aperture defined from the \cii\ velocity-integrated (moment-0) map, constructed by collapsing the cube over $-262$\,km\,s$^{-1}$ to $316$\,km\,s$^{-1}$ (for a \cii\ velocity range over ${\pm}\,3\sigma_{[\rm C\,II]}$ around the mean velocity). 
The \cii\ emission is robustly detected, with a peak significance of 13.8$\sigma$ in the moment-0 map and an integrated line significance of 13.5$\sigma$ in the extracted spectrum. 
We also extracted separate \cii\ spectra for the head and tail using the apertures adopted in the component photometry analysis (Fig.~\ref{fig3}a). 
A position-velocity diagram extracted along the major axis shows a kinematically connected gaseous structure extending from the head toward the knot, with evidence for multiple downstream gaseous components (Extended Data Fig.\,\ref{exfig6}). 
The moment-1 map was produced with noise pixels masked based on the same masking technique used in the data cube deconvolution process (Fig.\,\ref{fig2}b). 
The \cii\ emission extends across both the stellar head and the tail, with a secondary peak at the knot position kinematically consistent with the tail, indicating that the head, tail, and knot in the {\it JWST} data belong to the same gaseous structure. 

The \nii\ emission is detected at lower significance. 
To maximize the signal-to-noise ratio, we constructed the \nii\ moment-0 map by collapsing the cube over ${\pm}\,1\sigma_{[\rm N\,II]}$ around the mean velocity, corresponding to $-205$\,km\,s$^{-1}$ to $-99$\,km\,s$^{-1}$. 
The resulting map reveals an extended northern component and an unresolved southern component. 
We extracted the \nii\ spectra from the Kron aperture for the northern component and from the peak pixel for the southern component (Extended Data Fig.~\ref{exfig5}). 
The \nii\ line is detected at 4.4$\sigma$ and 3.2$\sigma$ in the extracted spectra, and at 4.4$\sigma$ and 3.9$\sigma$ peak significance in the moment-0 map, for the northern and southern components, respectively.

We measured their line intensities within ${\pm}\,3\sigma_{\rm line}$ of the fitted line centres and derived the corresponding line luminosities. 
The quoted uncertainties include both the statistical error and a 10\% absolute flux calibration uncertainty for ALMA Band 7 (Extended Data Table~\ref{extable4}). 
Under the standard $L_{\rm [C\,II]}$-to-$M_{\rm mol}$ conversion factor, $\alpha_{\rm [C\,II]}\,{=}\,30\,\,M_\odot\,L_\odot^{-1}$ \cite{Zanella2018}, the \cii\ luminosities imply molecular gas masses of $(3.5\,{\pm}\,0.5)\,{\times}\,10^{10}\,M_\odot$, $(1.4\,{\pm}\,0.2)\,{\times}\,10^{10}\,M_\odot$, and $(2.0\,{\pm}\,0.3)\,{\times}\,10^{10}\,M_\odot$ for the whole system, the head, and the tail, respectively.

For CO(2--1) and CO(4--3) observations at low frequency, the cosmic microwave background (CMB) effect is no longer negligible and must be taken into account. 
The elevated CMB temperature at high redshift could greatly reduce the contrast between emission and the CMB at a long wavelength \cite{Zhang2016}. 
The suppressing factor at a given frequency $\nu$ can be expressed as 
\begin{equation}
    f_{\nu}(T_{\rm ex},z)=\frac{J_\nu(T_{\rm ex}) - J_\nu(T_{\rm CMB}(z))}{J_\nu(T_{\rm ex}) - J_\nu(T_{\rm CMB}(0))},
\end{equation}
where $J_\nu(T) = (h\nu/k)/[\exp(h\nu/kT)-1]$ is the Planck radiation temperature at a given frequency $\nu$ and temperature $T$, $T_{\rm ex}$ is the gas excitation temperature, and $T_{\rm CMB}(z)$ is the CMB temperature at a given redshift. 
Adopting a fiducial gas excitation temperature $T_{\rm ex}=25\rm\,K$, the suppression factor would be $f_{\rm CO(2-1)}\approx0.52$ and $f_{\rm CO(4-3)}\approx0.61$ for CO(2--1) and CO(4--3), respectively. 

The ALMA Band-1 data show a CO(2--1) detection in C26 at $3.1\sigma$, with a spectral profile and spatial distribution broadly consistent with the \cii\ emission (Fig.~\ref{fig3}b). 
The observed integrated line flux is $S_{\rm CO(2-1)}\Delta v = 0.085\pm0.028\,\rm Jy\,km\,s^{-1}$ with the absolute calibration uncertainty included, corresponding to $L'_{\rm CO(2-1)} = (3.0\pm1.0)\times10^{10}\,\rm K\,km\,s^{-1}\,pc^2$ after CMB correction. 
Assuming a typical DSFG excitation condition with a CO(2--1)-to-CO(1--0) ratio of $r_{21}=0.85$ and a CO-to-H$_2$ conversion factor of $\alpha_{\rm CO}=0.8\,M_\odot\,(\rm K\,km\,s^{-1}\,pc^2)^{-1}$ \cite{Carilli2013,Aravena2016}, we infer a molecular gas mass of $M^{\rm CO(2-1)}_{\rm mol} = (2.8\pm0.9)\times10^{10}\,M_\odot$. 
This value is somewhat lower than the \cii-based estimate, but remains consistent within uncertainties. 
The difference may reflect the uncertainties in the adopted excitation and conversion factors or a neutral-gas component traced by \cii\ that is not fully traced by CO. 
By contrast, despite 16 hours of on-source integration in ALMA Band-3, the CO(4--3) line is not detected in C26 (Fig.~\ref{fig3}b). 
Using the same velocity interval and extraction aperture adopted for CO(2--1), we derive a 3$\sigma$ upper limit of $S_{\rm CO(4-3)}\Delta v < 0.051\,\rm Jy\,km\,s^{-1}$. 
Combined with the CO(2--1) detection, this implies a low excitation ratio of $r_{42}=L'_{\rm CO(4-3)}/L'_{\rm CO(2-1)}<0.13$, suggesting that the molecular gas in C26 is predominantly in a low-excitation state. 

The dust continuum is detected at 850\,$\mu$m with a peak significance of 4.6$\sigma$, but remains undetected at 3.2\,mm. 
The 850-$\mu$m flux density measured within the Kron aperture is $0.37\,{\pm}\,0.08$\,mJy, where the uncertainty includes both the statistical error and a 10\% absolute flux calibration error. 
The 3$\sigma$ upper limit on the 3.2-mm continuum flux density is ${<}\,8.9\,\rm\mu Jy$. 
We use the following formula to estimate its gas mass from the 850-$\mu$m continuum emission in units of $10^{10}\,M_\odot$ \cite{Scoville2016}
\begin{equation}
    M_{\rm mol}=1.78\,S_{\rm obs}(1+z)^{-(3+\beta)}\left(\frac{\nu_{850}}{\nu_{\mathrm{obs}}}\right)^{\beta+2}D_{\rm L}^2 \frac{\Gamma_{\rm RJ}}{\Gamma_0},
\end{equation}
where $S_{\rm obs}$ is the observed continuum flux density in unit of mJy, $\beta$ is the dust emissivity index, $\nu_{850}$ and $\nu_{\rm obs}$ are the frequencies at $850\,\mu\rm m$ and at the observed wavelength, $D_{\rm L}$ is the luminosity distance in Gpc, and $\Gamma_{\rm RJ}$ and $\Gamma_0$ are the correction factors for departures from strict $\nu^2$ of the Rayleigh-Jeans tail at observed and rest-frame wavelengths. 
With the standard $\beta=1.8$ and $T_{\rm dust}=25\rm\,K$, we use $850\rm\,\mu m$ continuum emission and derive $M^{\rm dust}_{\rm mol}\,{=}\,(2.4\,{\pm}\,0.5)\,{\times}\,10^{10}\,M_\odot$, broadly consistent with the molecular gas mass estimates inferred from \cii\ and CO(2--1). 

\subsection*{The origin of the multiphase tail}
We analyzed the morphology and geometry of C26 to investigate the origin of its elongated tail. 
Gravitational interactions can perturb both the stellar and gaseous components, whereas hydrodynamic stripping acts preferentially on the diffuse interstellar medium \cite{Boselli2022}. 
We therefore compared the stellar morphology, gas excitation, and multiphase-ISM structure of C26, to assess which mechanism most likely dominates the observed feature. 

\subsubsection*{Morphology, geometry and resolved stellar structure}

We first quantified the morphology and projected geometry of the stellar tail in C26 using the {\it JWST} imaging. 
The galaxy shows a one-sided, cometary stellar tail extending from a compact stellar head, a morphology that is less consistent with the more symmetric, two-sided features often produced by gravitational interactions. 
We used the apertures defined in the component photometry to identify the head and tail. 
We then measured the tail position angle from the major axis of the elongated stellar structure, using the second-order moments of the F444W light distribution. 
The tail position angle is 75.2\,deg, consistent with the cluster-centric axis defined by the tSZ signal (PA$\,{=}\,73.4\pm4.9$\,deg) within $1\sigma$ uncertainty, and nearly perpendicular to the direction of the nearest neighboring galaxy (PA$\,{=}\,$1.5\,deg). 
The one-sided tail, aligned with the cluster-centric axis, is therefore more naturally explained by its motion through the cluster environment than by a direct interaction with the neighboring galaxy. 

We next examined the resolved stellar structure of the tail to further test a tidal interpretation. 
While the tail and the star-forming knot are clearly detected in both the F444W and F1000W images, the higher-resolution F200W image resolves the stellar emission into multiple clumps. 
These clumps are unresolved or marginally resolved, with half-light radii of $R_{e}\,{\leq}\,0.1\,$arcsec, corresponding to physical sizes of $R_e\,{\leq}\,0.7\rm\,kpc$. 
We then matched the PSFs of the F200W and F444W images and constructed a spatially-resolved $\rm F200W\,{-}\,F444W$ color map with noisy pixels (${<}\,3$ times the rms) masked to examine the colour gradient across C26 (Extended Data Fig.~\ref{exfig3}). 
Relative to the stellar head, the tail is more than 1\,mag bluer overall, except for the star-forming knot, which is on average only ${\approx}\,0.3\rm\,mag$ bluer. 
The redder color of the knot may reflect additional dust attenuation from obscured star formation suggested by the SED fitting, as it coincides with the enhanced \cii\ emission and tentative $2\sigma$ excess in the 850-$\mu$m continuum. 
The clumpy morphology and color gradients within the tail are consistent with recent star formation in the stripped material. 
Such a young, clumpy, one-sided stellar structure therefore disfavors a simple tidal extension of pre-existing stellar light. 

\subsubsection*{Star formation and molecular-gas excitation}

The star-forming and molecular-gas properties provide an additional test of an interaction-driven origin. 
Although C26 contains a massive cold-gas reservoir, its star-formation activity remains modest compared with that expected for strongly interaction-driven starburst. 
The CO(4--3) non-detection further implies a very low molecular-gas excitation ratio ($r_{42}<0.13$), indicating that the gas is not dominated by dense, warm molecular clouds with active star formation. 
Instead, the molecular reservoir appears sub-thermally excited, broadly consistent with stripped gas in the tails of low-redshift jellyfish galaxies \cite{Moretti2018,Moretti2020}. 
These properties do not uniquely distinguish RPS from past tidal encounter or high-velocity flyby, but argue against a strongly interaction-driven starburst as the dominant mechanism shaping the tail. 

\subsubsection*{Multiphase ISM structure}
We next investigated the multiphase structure of the tail to test whether it is consistent with hydrodynamic stripping.

The relative distributions of the cold gas and dust support this interpretation. 
Both the \cii\ and 850-$\mu$m continuum emission are offset from the stellar emission peak of the head by $\Delta\alpha_{\rm [C\,II]}\,{=}\,0.84\,{\pm}\,0.04\,\,(0.87\,{\pm}\,0.04)$\,arcsec and $\Delta\alpha_{\rm dust} \,{=}\,0.55\,{\pm}\,0.05\,\,(0.58\,{\pm}\,0.05)$\,arcsec in the F444W (F200W) band, where both astrometric and positional uncertainties are included \cite{Ivison2007}. 
The corresponding projected physical offsets are $5.8\,{\pm}\,0.3\,\,(6.0\,{\pm}\,0.3)$ \,kpc and $3.8\,{\pm}\,0.3\,\,(4.0\,{\pm}\,0.3)$\,kpc for \cii\ and dust emission, respectively. 
The smaller offset of the dust emission is consistent with phase-dependent stripping in a hydrodynamic interaction, in which the more extended gas traced by \cii\ is displaced more efficiently than the denser material traced by the dust continuum.

The ionized gas distribution provides additional support that the displaced material is interacting with the surrounding medium. 
The \nii\ emission does not follow the central \cii\ ridge, but is instead enhanced on both flanks of the tail. 
Its velocities are also consistent with those of the \cii\ emission extracted at the corresponding positions in the tail. 
This flank-brightened \nii\ morphology indicates ionized interface layers around the displaced cold gas, as expected where the tail is exposed to and mixed with the surrounding hot medium. 
These multiphase features are consistent with stripped gas interacting with the surrounding hot medium \cite{Poggianti2019,Sun2021}. 

\subsection*{Required stripping conditions}
To assess whether such a stripping event is physically plausible at $z\,{=}\,4.3$, we estimated the conditions required for RPS in C26. 
The minimum external pressure needed to overcome the galaxy’s restoring force can be written as 
\begin{equation}
    P_{\rm min}(r)=2\pi\,G\,\Sigma_\star(r)\,\Sigma_{\rm gas}(r),
\end{equation}
where $\Sigma_\star(r)$ and $\Sigma_{\rm gas}(r)$ are the stellar and gas surface-density profiles as a function of galactocentric radius \cite{Gunn1972}. 
While the expression assumes face-on stripping, hydrodynamic simulations show that the stripping radius depends only weakly on inclination for $i\lesssim60^\circ$, with inclination effects arising primarily for near edge-on motion \cite{Roediger2006}. 

For simplicity, we assumed an exponential disc profile for the surface density \cite{Boselli2006,Bellhouse2017}
\begin{equation}
    \Sigma(r)=\left(\frac{M}{2\pi\,r_d^2}\right)e^{-r/r_d},
\end{equation}
where $M$ is the component mass and $r_d$ is the disc scale length. 
The stellar mass was obtained from the SED fitting ($M_\star\,{=}\,2.2\,{\times}\,10^{10}\,M_\odot$). 
As the effective dust-to-gas ratio may be substantially lower in C26 and the molecular gas mass inferred from \cii\ broadly agrees with the CO(2--1) measurement, we considered a total gas mass of $M_{\rm gas}\,{\approx}\,M_{\rm mol}^{\rm [CII]}\,{=}\,3.5\,{\times}\,10^{10}\,M_\odot$. 
We estimated the stellar scale length to be $r_{d,\star}\,{=}\,1.2$\,kpc from the F444W image of the stellar head. 
For the gas, we assumed $r_{d,\rm gas}\,{=}\,1.7\,r_{d,\star}\,{=}\,2.0\rm\,kpc$ \cite{Boselli2006}. 
These assumptions give 
\begin{equation}
    P_{\rm min}(r)=(6.2\times10^{-9}\rm\,N/m^2)\times e^{-(r/r_{d,\star}+r/r_{d,\rm gas})}
\end{equation}

The ram pressure can be expressed as \cite{Gunn1972}
\begin{equation}
    P_{\rm ram}=\rho_{\rm ICM}\Delta v_{\rm 3D}^2,
\end{equation}
where $\rho_{\rm ICM}$ is the ambient ICM density and $\Delta v_{\rm 3D}$ is the three-dimensional velocity of the galaxy relative to the surrounding medium. 
Because the line-of-sight velocity of C26 is close to the systemic velocity of the protocluster, the true three-dimensional differential velocity is uncertain. 
Based on the projected 2D velocity dispersion of the protocluster core \cite{Hill2020,Sulzenauer2025}, we considered three illustrative cases: 500\,km\,s$^{-1}$, 1000\,km\,s$^{-1}$, and 1500\,km\,s$^{-1}$, which represent 1$\sigma$, 2$\sigma$, and $3\sigma$ of the velocity distribution without considering ICM bulk motion. 

By combining Eqs.\,(4) and (5), we obtained the initial stripping radius as a function of ambient ICM density for the different velocity and gas-mass assumptions shown in Fig.~\ref{fig4}. 
As a reference value, we indicate the gas density corresponding to 15\% of the characteristic overdensity $\rho_{500}$ at $z\,{=}\,4.3$. 
For plausible ICM densities at this redshift, substantial gas displacement can be achieved. 
However, the exact stripping radius depends sensitively on the adopted gas mass and the true galaxy velocity relative to the surrounding medium. 
Since the critical density at $z\,{=}\,4.3$ is about 50 times larger than at $z\,{=}\,0$, the ambient density in a hot protocluster core could also be substantially higher \cite{Boselli2022}. 
In addition, a dynamically disturbed ICM may contain locally compressed regions and exhibit turbulence, shocks, or bulk motions, which would increase both the ambient gas density and the effective differential velocity between the galaxy and the surrounding medium \cite{Li2018,Ge2026}. 
Together, these effects would naturally enhance the ram pressure acting on infalling galaxies, and may help explain why such strong stripping can be observed in SPT2349$-$56 at this early epoch.

\subsection*{Survival of the stripped gas}
To test whether the displaced cold gas can survive long enough to form the observed tail, we estimate the characteristic timescale for hydrodynamic disruption, including Kelvin-Helmholtz instabilities, using the cloud-crushing timescale \cite{Klein1994}
\begin{equation}
    t_{\rm cc} \approx \sqrt{\chi}\frac{r_{\rm cl}}{\Delta v},
\end{equation}
where $\chi=\rho_{\rm cl}/\rho_{\rm ICM}$ is the density contrast between the cold gas and ambient ICM; $r_{\rm cl}$ is the characteristic cloud radius; $\Delta v$ is the relative velocity between the cloud and ambient gas. 
We adopt a fiducial cold gas density for the \cii-emitting gas of $n_{\rm H_2}=10^3\,\rm cm^{-3}$ (or $\rho_{\rm cl}\approx4.7\times10^{-18}\rm\,kg\,m^{-3}$), which represents the internal density of gas cloud rather than the volume-averaged density of the full tail. 
Assuming an ambient ICM density $\rho_{\rm ICM}=0.15\rho_{\rm 500}\approx3.0\times10^{-23}\rm\,kg\,m^{-3}$, and a fiducial cloud size $r_{\rm cl}\approx 0.7\rm\,kpc$, we obtain $t_{\rm cc}\approx550, 270,$ and $180\rm\,Myr$ for $\Delta v=500, 1000,$ and $1500\rm\,km/s$, respectively. 

Meanwhile, the characteristic formation timescale required to produce the observed tail \cite{Jachym2017} can be estimated as 
\begin{equation}
    t_{\rm form}\sim\frac{L_{\rm tail}}{\Delta v},
\end{equation}
where $L_{\rm tail}\approx20\rm\,kpc$ is the length of the observed tail. 
For the same velocities, the corresponding $t_{\rm form}$ would be 39, 19, 13\,Myr. 
The resulting cloud-crushing times are more than an order of magnitude longer than the corresponding formation times ($t_{\rm cc}/t_{\rm form}\sim\sqrt{\chi}\,r_{\rm cl}/L_{\rm tail}\approx14$). 
Therefore, under these fiducial assumptions, hydrodynamic disruption is not expected to destroy the cold-gas structures before the observed tail can form.
\clearpage
\setcounter{figure}{0}
\makeatletter 
\renewcommand{\thefigure}{\@arabic\c@figure}
\renewcommand{\thetable}{\@arabic\c@table}
\renewcommand{\figurename}{Extended Data Fig.}
\renewcommand{\tablename}{Extended Data Table}
\makeatother

\begin{figure}
    \centering
    \includegraphics[width=0.98\linewidth]{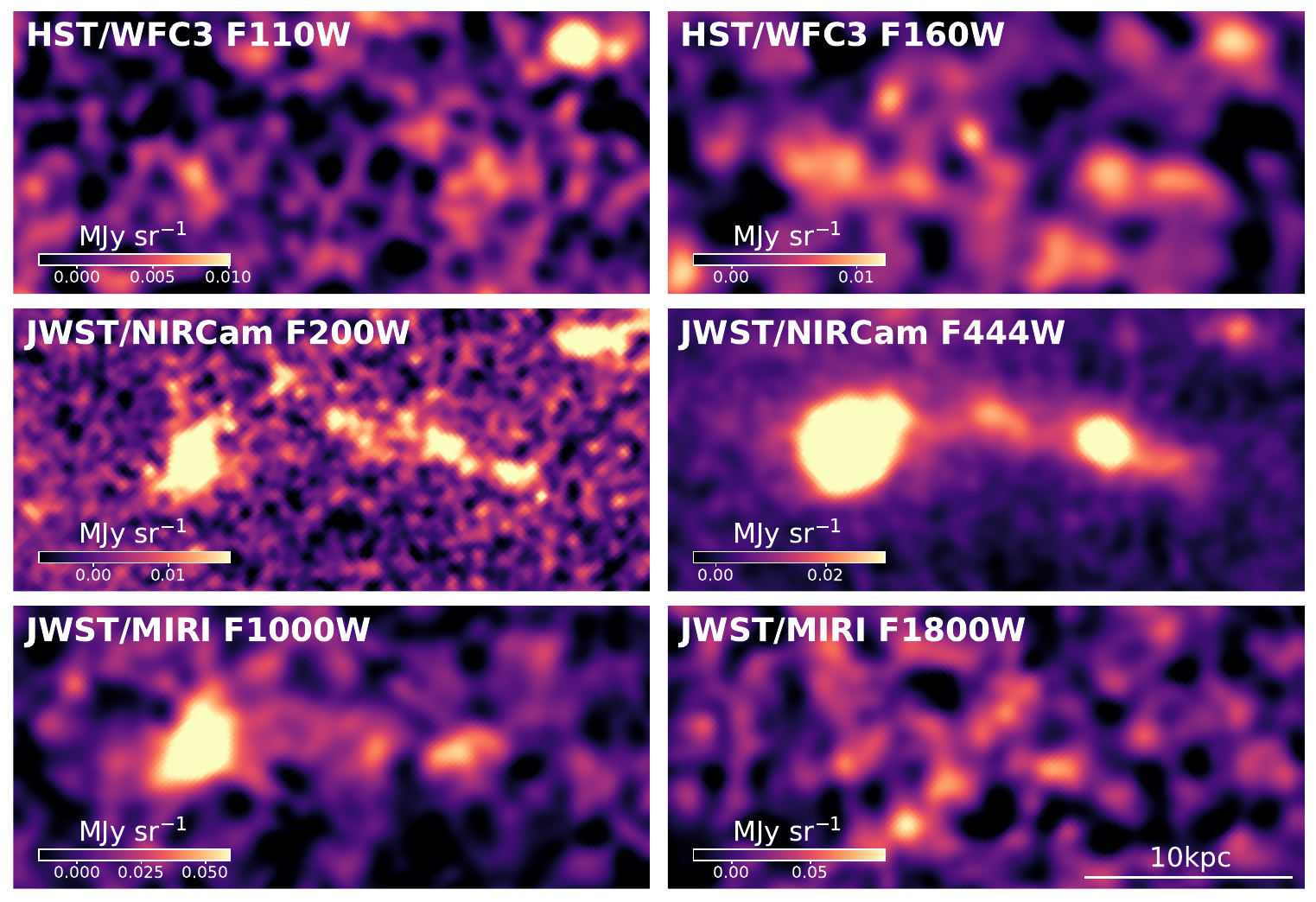}
    \caption{{\bf {\it HST} and {\it JWST} imaging of C26 across six bands.}\\
    Cutout images of C26 in {\it HST}/WFC3 F110W and F160W, {\it JWST}/NIRCam F200W and F444W, and {\it JWST}/MIRI F1000W and F1800W sampled on the NIRCam F200W grid at native resolution. 
    The white bar in the lower-right panel indicates a projected scale of 10 kpc.
    }
    \label{exfig1}
\end{figure} 

\begin{figure}
    \centering
    \includegraphics[width=0.98\linewidth]{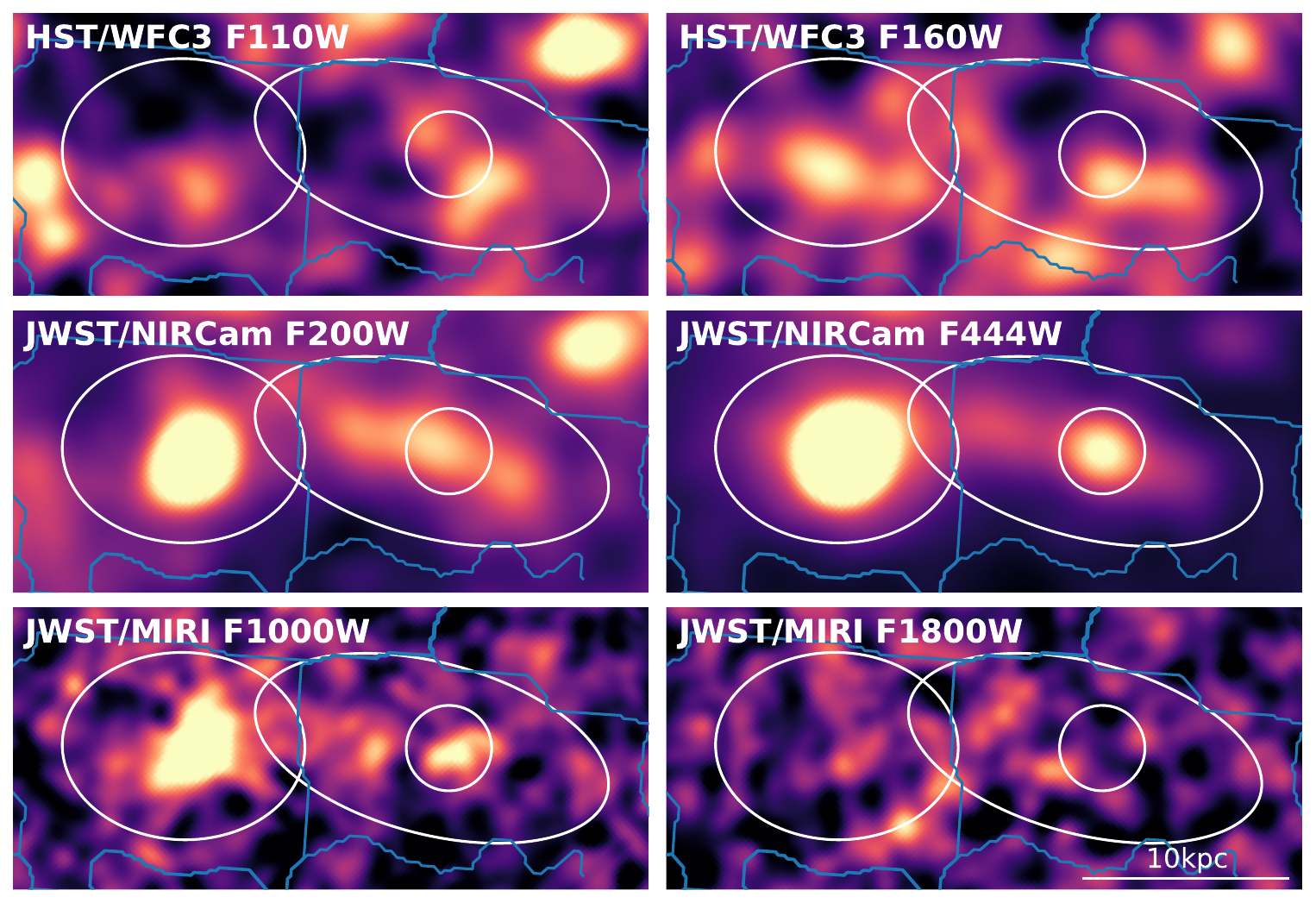}
    \caption{{\bf Image cutouts and apertures for component photometry}\\
    Cutouts of C26 in the {\it HST}/WFC3 F110W and F160W bands, the {\it JWST}/NIRCam F200W and F444W bands, and the {\it JWST}/MIRI F1000W and F1800W bands. 
    All images except F1800W are PSF-matched to a common resolution. 
    The F1800W image is shown at its native resolution. 
    White ellipses show apertures for the component photometry of the head, tail, and knot. 
    Blue contours indicate the segmentation map used to mask nearby sources to prevent contamination. 
    }
    \label{exfig2}
\end{figure}

\begin{figure}
    \centering
    \includegraphics[width=0.98\linewidth]{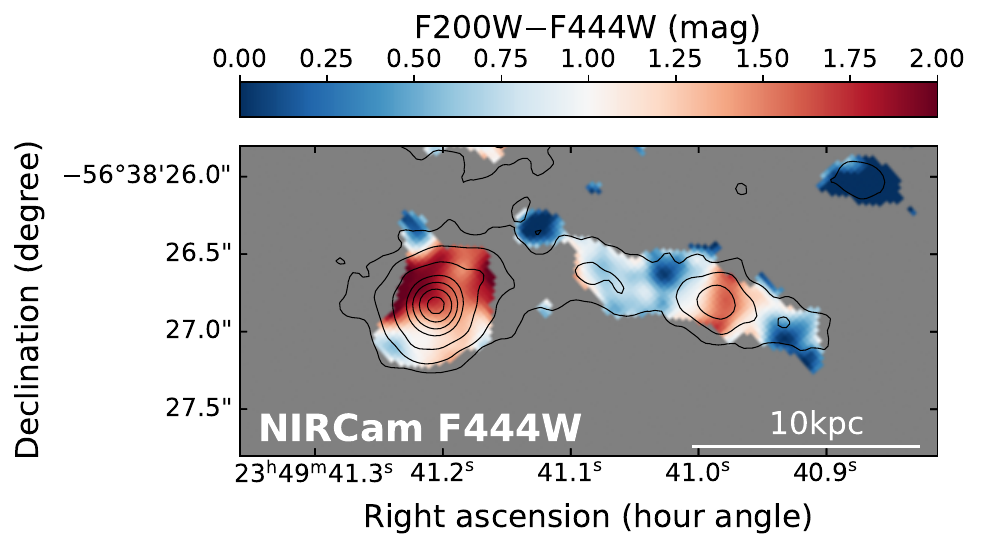}
    \caption{{\bf PSF-matched F200W$\,{-}\,$F444W color map of C26}\\
    Spatially-resolved F200W$\,{-}\,$F444W color map of C26 constructed from the PSF-matched NIRCam images. 
    Noisy pixels below $3\sigma$ in either F200W or F444W image have been masked for clarity. 
    Black contours show the F444W emission at [5, 10, 20, 40, 60, 90, 130] times the local rms. 
    The stellar tail is bluer than the stellar head overall, while the compact knot embedded in the tail is redder than the surrounding diffuse emission, consistent with the additional dust attenuation indicated by the SED fitting and concentration of the \cii\ emission. 
    }
    \label{exfig3}
\end{figure} 

\begin{figure}
    \centering
    \includegraphics[width=0.98\linewidth]{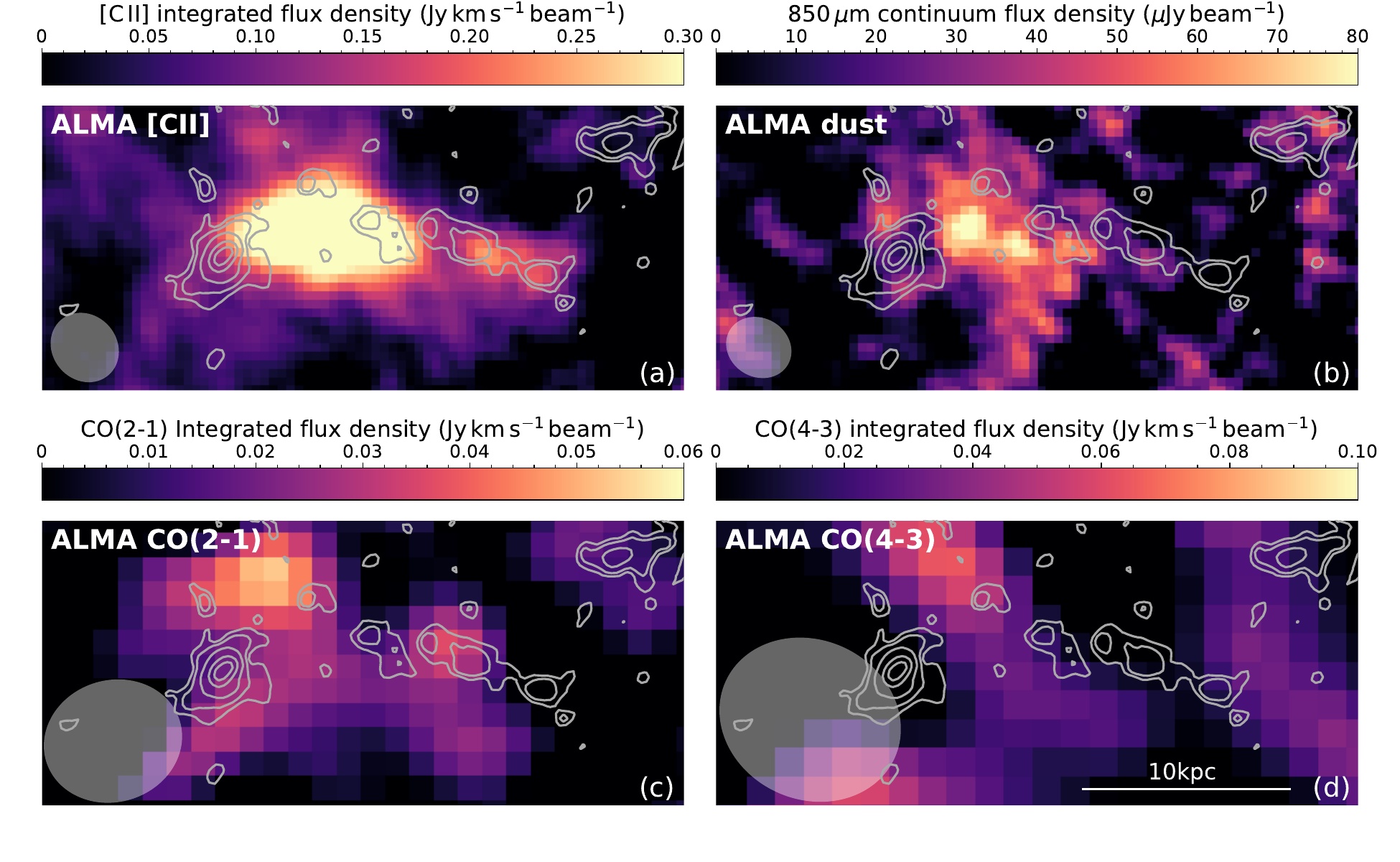}
    \caption{{\bf ALMA moment-0 maps of cold ISM tracers in C26.}\\
    Moment-0 maps of (a) \cii, (b) 850-$\mu$m continuum, (c) CO(2--1), and (d) CO(4--3) line emission in C26. 
    White contours show [3, 5, 10, 20, 30] times the local rms of the NIRCam F200W image for reference. 
    White ellipses in the lower-left corners indicate the synthesized beams. 
    The \cii\ emission traces the extended cold-gas reservoir, while the dust and CO lines probe denser components. 
    }
    \label{exfig4}
\end{figure} 

\begin{figure}
    \centering
    \includegraphics[width=0.95\linewidth]{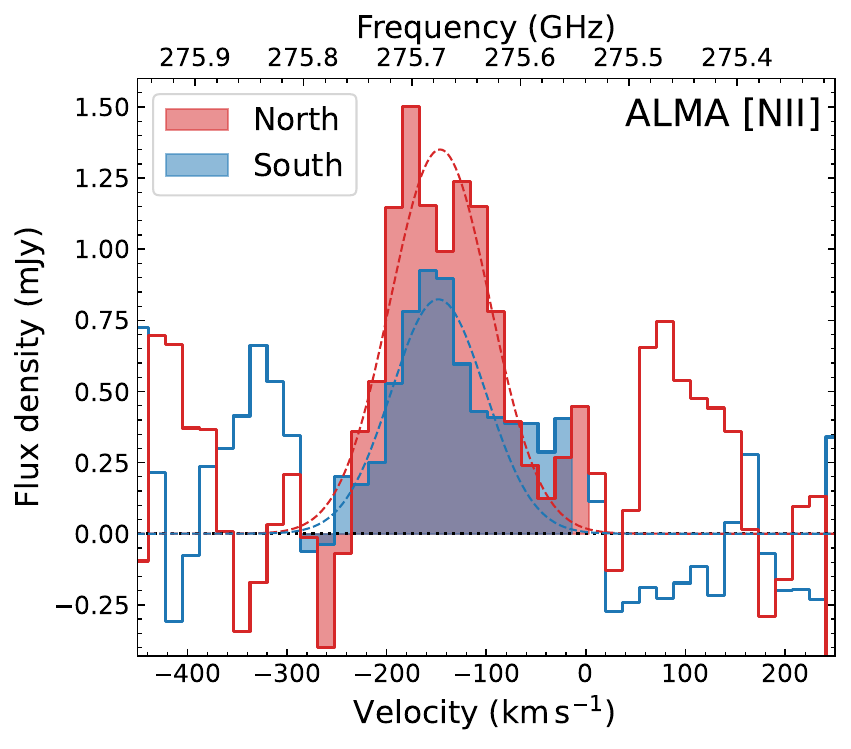}
    \caption{{\bf \nii\ spectra of the northern and southern tail components.}\\
    ALMA \nii\ spectra extracted from the northern (red) and southern (blue) components identified in the \nii\ moment-0 map. 
    The dashed lines are the best-fit Gaussian profiles for determining the velocity ranges for velocity-integrated intensity calculations, which are indicated as the shaded regions. 
    Low-level features near $+100\rm\,km\,s^{-1}$ in the northern spectrum and $-300\,\rm km\,s^{-1}$ in the southern spectrum remain below our $3\sigma$ detection threshold and do not show spatially coherent emission in channel maps, which are therefore not treated as significant part of \nii\ emission.
    }
    \label{exfig5}
\end{figure} 

\begin{figure}
    \centering
    \includegraphics[width=0.9\linewidth]{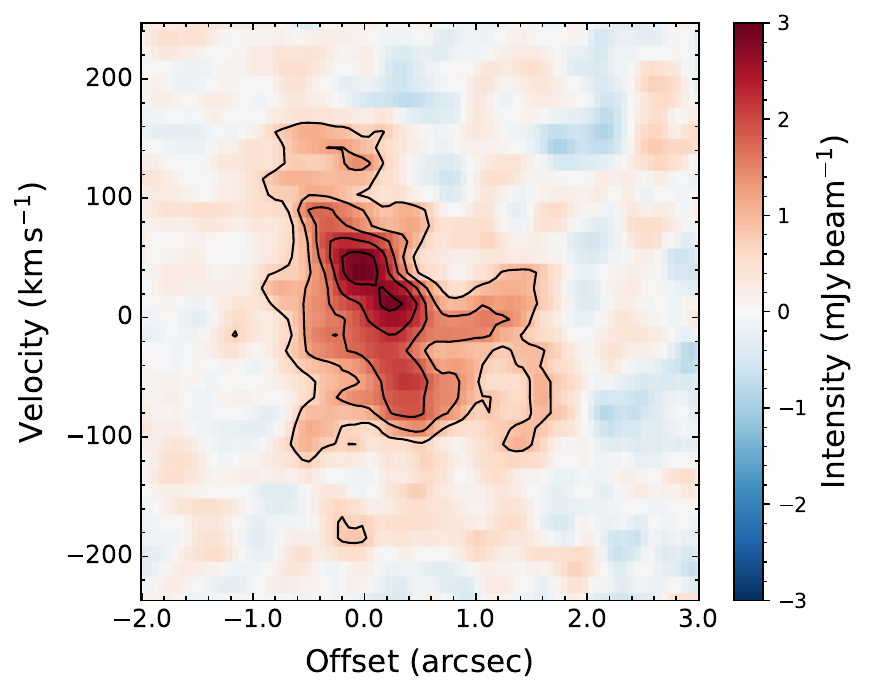}
    \caption{{\bf Position-velocity diagram of the \cii\ tail in C26}\\
    Position-velocity diagram of the ALMA \cii\ emission in C26, extracted along the major axis. 
    The color scale shows the primary-beam-corrected intensity, and black contours indicate [3, 5, 7, 9, 11, 13] times the local rms. The \cii\ emission remains kinematically connected from the stellar head to the downstream knot, supporting the interpretation that the head, tail, and knot belong to the same gaseous structure. 
    The diagram also shows evidence for multiple downstream gaseous components.
    }
    \label{exfig6}
\end{figure} 

\clearpage
 
\begin{table*}[]
    \centering
    \caption{\textbf{\textit{HST}} {\bf and} \textbf{\textit{JWST}} {\bf photometry}}    \label{extable2}
    \renewcommand{\arraystretch}{1.3}
    \begin{tabular}{cccc}
    \hline
    \hline
       Band  & Head & Tail & Knot\\ 
       & (nJy) & (nJy) & (nJy)  \\
       \hline
        {\it HST}/WFC3 F110W & $31\pm12$ &  $84\pm15$ & $29\pm10$ \\ 
        {\it HST}/WFC3 F160W & $143\pm21$ & $130\pm27$  & $34\pm16$\\ 
        {\it JWST}/NIRCam F200W & $254\pm18$ & $214\pm22$  & $74\pm12$\\ 
        {\it JWST}/NIRCam F444W & $1134\pm10$ & $515\pm12$ & $214\pm8$\\  
        {\it JWST}/MIRI F1000W & $1085\pm99$ & $609\pm126$ & $264\pm78$\\
        {\it JWST}/MIRI F1800W & $821\pm218$ & $664\pm248$ & $158\pm226$\\ 
        \hline
    \end{tabular}
\end{table*}

\begin{table*}[]
    \centering
    \caption{{\bf Physical properties of C26 from the MAGPHYS SED fitting.}}    \renewcommand{\arraystretch}{1.6}  
    \label{extable3}
    \begin{tabular}{ccccc}
    \hline
    \hline
       Component  & $M_{\star}$  & $A_{\rm v}$ & SFR  & sSFR \\
       & ($10^9\rm\,M_\odot$) & (mag) & ($\rm M_\odot\,yr^{-1}$) & ($10^{-9}$\,yr$^{-1}$)\\
       \hline
        Head & $21.9^{+7.0}_{-4.5}$ & $1.41^{+0.35}_{-0.10}$  & $30.3^{+17.2}_{-8.4}$ & $1.33^{+0.55}_{-0.49}$ \\
        Tail & $5.9^{+1.6}_{-1.3}$ & $0.94^{+0.40}_{-0.35}$  & $18.3^{+13.9}_{-6.9}$ & $2.99^{+2.97}_{-1.10}$ \\
        Knot & $3.2^{+0.8}_{-0.8}$ & $1.11^{+0.35}_{-0.43}$  & $7.6^{+4.7}_{-3.3}$ & $2.37^{+1.85}_{-1.04}$ \\
        \hline
    \end{tabular}
\end{table*}

\begin{table*}[]
    \centering
    \caption{{\bf Summary of the ALMA datasets used in this work.} }   \label{extable1}
    \makebox[\textwidth][c]{
    \hspace*{-0.6cm}
    \renewcommand{\arraystretch}{1.3}
    \begin{tabular}{ccccccccc}
    \hline
    \hline
        Program ID & Mosaic & $L_{\rm base}$ & $t_{\rm source}$ & Emission & Frequency & $\sigma$ & Channel width & Beam\\
       & & (m) & (min) & & (GHz) & ($\mu$Jy\,beam$^{-1}$) & (km\,s$^{-1}$) & (arcsec$^2$)\\
       \hline
        2016.1.00236.T & No & 15--460 & 14 & \multirow{2}{*}{Dust} & \multirow{2}{*}{352.41} & \multirow{2}{*}{11} & \multirow{2}{*}{...} & \multirow{2}{*}{$0.47\,{\times}\,0.41$} \\
        2017.1.00273.S & 3-point & 15--484 & 173\\
        2018.1.00058.S & 6-point & 15--1398 & 239 & \\
        \multirow{2}{*}{2021.1.01063.S} & 7-point & 15--680 & 318 & \cii & 358.40 & 128 & 13 & $0.51\,{\times}\,0.45$ \\
        & No & 15--500 & 18  \\
        \hline
        2016.1.00236.T & No & 15--460 & 22 & \multirow{2}{*}{\nii} & \multirow{2}{*}{275.54} & \multirow{2}{*}{126} & \multirow{2}{*}{17} & \multirow{2}{*}{$0.74\,{\times}\,0.64$} \\
        2017.1.00273.S & No & 15--500 & 84 \\
        \hline
        2023.1.00895.S & No & 15--2231 & 440 & CO(2--1) & 43.47 & 77 & 54 & $0.98\,{\times}\,0.85$ \\
        \hline
        2015.1.01543.T & No & 15--704 & 47 &  \\
        2017.1.00273.S & No &15--1231 & 155 & CO(4--3) & 86.94 & 58 & 54 & $1.30\,{\times}\,1.12$\\
        2023.1.00124.S & No & 15--783 & 782 \\
        \hline
        2015.1.01543.T & No & 15--704 & 47 & \multirow{4}{*}{Dust} & \multirow{4}{*}{95.06} & \multirow{4}{*}{2.6} & \multirow{4}{*}{...} & \multirow{4}{*}{$1.31\,{\times}\,1.16$} \\
        2017.1.00273.S & No & 15--1231 & 241 \\
        2022.1.00495.S & No & 15--368 & 346\\
        2023.1.00124.S & No & 15--783 & 782 \\
        \hline
    \end{tabular}
    }

\end{table*}

\begin{table*}[]
    \centering
    \caption{{\bf Line intensities and molecular gas estimates.}}    \label{extable4}
    \renewcommand{\arraystretch}{1.3} 
    \begin{tabular}{ccccc}
    \hline
    \hline
        Line & Component & $S_{\rm line}\Delta v$ & $L_{\rm line}$ & $M_{\rm mol}$ \\
        & & (Jy\,km\,s$^{-1}$) & ($10^7\rm\,L_\odot$)  & ($10^{9}\rm\,M_\odot$)  \\
        \hline
        \cii & Full & $1.96\pm0.29$ & $116.6\pm17.0$ & $35.0\pm5.1$ \\
        \cii & head & $0.80\pm0.12$ & $47.5\pm6.9$ & $14.2\pm2.1$\\
        \cii & tail & $1.09\pm0.15$ & $64.9\pm8.9$ & $19.5\pm2.7$\\
        \nii & north & $0.17\pm0.04$ & $8.0\pm1.9$ & ... \\
        \nii & south & $0.12\pm0.03$ & $5.4\pm1.5$ & ... \\
        CO(2--1) & Full & $0.085\pm0.028$ & $1.2\pm0.4$ & $28.2\pm9.3$ \\
        CO(4--3) & Full & $<0.051$ & $<1.2$ & $<5.1$ \\
        \hline
    \end{tabular}
\end{table*} 
\bmhead{Correspondence statement}
Correspondence and requests for materials should be addressed to D.Z. (\href{mailto:dzhou.astro@gmail.com}{dzhou.astro@gmail.com}).
\bmhead{Acknowledgments} 
We are grateful to Leslie Sage for his valuable guidance and thoughtful feedback, which greatly improved the clarity and presentation of the manuscript. 
We thank Toby Brown, Chong Ge, Daizhong Liu, Masao Mori, Allison Noble, Douglas Scott, Antony Stark, Daniel Wang, Zhi-Yu Zhang for useful discussions on this work. 
This paper makes use of the following ALMA data: ADS/JAO.ALMA\#2015.1.01543.T, ADS/JAO.ALMA\#2016.1.00236.T, ADS/JAO.ALMA\#2017.1.00273.S, ADS/JAO.ALMA\#2018.1.00058.S, ADS/JAO.ALMA\#2021.1.01063.S, ADS/JAO.ALMA\#2022.1.00495.S, ADS/JAO.ALMA\#2023.1.00124.S, ADS/JAO.ALMA\#2023.1.00895.S. 
ALMA is a partnership of ESO (representing its member states), NSF (USA) and NINS (Japan), together with NRC (Canada), NSTC and ASIAA (Taiwan), and KASI (Republic of Korea), in cooperation with the Republic of Chile. The Joint ALMA Observatory is operated by ESO, AUI/NRAO and NAOJ. 
The National Radio Astronomy Observatory is a facility of the National Science Foundation operated under cooperative agreement by Associated Universities, Inc. 
This research used the Canadian Advanced Network For Astronomy Research (CANFAR) operated in partnership by the Canadian Astronomy Data Centre and The Digital Research Alliance of Canada with support from the National Research Council of Canada the Canadian Space Agency, CANARIE and the Canadian Foundation for Innovation. 
This research is based on observations made with the NASA/ESA Hubble Space Telescope obtained from the Space Telescope Science Institute, which is operated by the Association of Universities for Research in Astronomy, Inc., under NASA contract NAS 5-26555. These observations are associated with program \#15701. 
This work is based [in part] on observations made with the NASA/ESA/CSA James Webb Space Telescope. The data were obtained from the Mikulski Archive for Space Telescopes at the Space Telescope Science Institute, which is operated by the Association of Universities for Research in Astronomy, Inc., under NASA contract NAS 5-03127 for JWST. These observations are associated with program \#06669. 
The SPT is supported by the NSF through grant OPP-1852617. 
This work was partially supported by the Center for AstroPhysical Surveys (CAPS) at the National Center for Supercomputing Applications (NCSA), University of Illinois Urbana-Champaign. 
D.Z., S.C.C., R.H., and G.C.P.W. acknowledge support from NSERC-6740. 
This research was supported in part by grant NSF PHY-2309135 to the Kavli Institute for Theoretical Physics (KITP) during the fall 2025 residence (S.C.C.). 
M.A. is supported by FONDECYT grant number 1252054, and gratefully acknowledges support from ANID Basal Project FB210003 and ANID MILENIO NCN2024\_112.

\bmhead{Author contribution}
D.Z. reduced and analyzed the data, interpreted the results, produced the figures, and drafted the manuscript. 
S.C.C. supervised the projects. 
D.Z., S.C.C., M.A., R.D., A.H.G., R.H., N.L., M.A.M., A.M., N.Z., K.A.P., V.R.P., M.S., J.S.S., N.S., J.D.V., D.V., G.C.P.W., and A.W. contributed substantially to discussing the results and preparing the manuscript. 
\section*{Declarations}

\begin{itemize}
\item {\bf Data availability}: 
The ALMA (2015.1.01543.T, 2016.1.00236.T, 2017.1.00273.S, 2018.1.00058.S, 2021.1.01063.S, 2022.1.00495.S, 2023.1.00124.S, 
2023.1.00895.S), {\it HST} ({\it HST}-GO-15701), and {\it JWST} ({\it JWST}-GO-06669) data used in this work are publicly available on the ALMA science archive (https://almascience.nrao.edu/aq/) and MAST service (https://mast.stsci.edu). 

\item {\bf Code availability}: 
No custom code central to the conclusions was used. 
Presented results can be reproduced using the following publicly available packages: 
\texttt{astropy} \cite{astropy2022}, 
\texttt{astroquery} \cite{astroquery}, 
\texttt{CASA} \cite{CASA}, 
\texttt{matplotlib} \cite{matplotlib}, 
\texttt{numpy} \cite{numpy}, 
\texttt{pandas} \cite{pandas}, 
\texttt{photutils} \cite{photutils},
\texttt{spectral\_cube} \cite{spectralcube}. 
\item {\bf Competing Interests:} The authors declare no competing interests. 
\end{itemize}

%%=============================================%%
%% For submissions to Nature Portfolio Journals %%
%% please use the heading `Extended Data'.   %%
%%=============================================%%

%%=============================================================%%
%% Sample for another appendix section			       %%
%%=============================================================%%

%% \section{Example of another appendix section}\label{secA2}%
%% Appendices may be used for helpful, supporting or essential material that would otherwise 
%% clutter, break up or be distracting to the text. Appendices can consist of sections, figures, 
%% tables and equations etc.

% \end{appendices}

%%===========================================================================================%%
%% If you are submitting to one of the Nature Portfolio journals, using the eJP submission   %%
%% system, please include the references within the manuscript file itself. You may do this  %%
%% by copying the reference list from your .bbl file, paste it into the main manuscript .tex %%
%% file, and delete the associated \verb+\bibliography+ commands.                            %%
%%===========================================================================================%%

\bibliography{sn-bibliography}% common bib file
%% if required, the content of .bbl file can be included here once bbl is generated
%%\input sn-article.bbl

\end{document}